\documentclass[12pt, a4paper, fleqn]{article}

\usepackage[nottoc]{tocbibind} 

\RequirePackage[l2tabu, orthodox]{nag}
\usepackage{silence}
\ActivateWarningFilters[pdftoc]

\usepackage[T1]{fontenc}
\usepackage[utf8]{inputenc}

\usepackage[top=20mm, bottom=20mm, left=25mm, right=25mm]{geometry}

\usepackage{array,longtable}
\usepackage{multirow}

\usepackage[table]{xcolor}

\usepackage{amsmath, amssymb, amsfonts, amsthm}
\usepackage{centernot}
\usepackage{bm}  
\usepackage{mathtools} 
\usepackage{graphicx}

\usepackage[colorlinks=true, linktoc=all, linkcolor=black, citecolor=red, urlcolor=blue, backref=page]{hyperref}

\usepackage{etoolbox}
\makeatletter
\patchcmd{\BR@backref}{\newblock}{\newblock(}{}{}
\patchcmd{\BR@backref}{\par}{)\par}{}{}
\makeatother

\numberwithin{equation}{section}
\usepackage{tikz}
\usetikzlibrary{calc,decorations.pathmorphing}
\usetikzlibrary{decorations.pathreplacing}
\usetikzlibrary{decorations.markings, arrows.meta}
\usetikzlibrary{shapes}
\usepackage{pgf}
\usetikzlibrary{calc}

\newcommand{\vvv}{
  \coordinate (base) at (0, .8);
  \node[fill, circle, minimum size = 2mm, inner sep = 0] at (0, 0) (a) {};
  \node[fill, circle, minimum size = 2mm, inner sep = 0] at (3, 0) (c) {};
  \node[fill, circle, minimum size = 2mm, inner sep = 0] at (1.5, 2.6) (b) {};
}
\newcommand{\tever}{
  \coordinate (base) at (0, .8);
  \node[draw, circle, red!70!black] at (-2, 0) (a) {};
  \node[draw, circle, red!70!black] at (2, 0) (c) {};
  \node[draw, circle, red!70!black] at (0, 3) (b) {};
}

\tikzset{
  loops/.style={line width = 2pt, blue, rounded corners=6pt, opacity=0.55}
}

\newcommand{\uJTL}{u\mathcal{J\!T\!\!L}}

\begin{document}

\section*{}
\begin{flushleft}
  {\bfseries\sffamily\Large 
    Three-point functions in critical loop models
    \vspace{1.5cm}
    \\
    \hrule height .6mm
  }
  \vspace{1.5cm}

  {\bfseries\sffamily 
    Jesper Lykke Jacobsen$^{1,2,3}$, Rongvoram Nivesvivat$^4$, Sylvain Ribault$^1$, Paul Roux$^{1,2}$
  }
  \vspace{4mm}

  {\textit{\noindent
      \\ \vspace{2mm}
      $^1$
      Institut de physique théorique, CEA, CNRS, 
      Université Paris-Saclay
      \\ 
      $^2$ Laboratoire de Physique de l’École Normale Supérieure, ENS, Université PSL, CNRS, Sorbonne Université, Université Paris Cité, F-75005 Paris, France
      \\
      $^3$ Sorbonne Université, École Normale Supérieure, CNRS,
      Laboratoire de Physique
      \\
      $^4$ New York University Abu Dhabi, United Arab Emirates
    }}

  \vspace{4mm}

  {\textit{E-mail:} \texttt{
      jesper.jacobsen@phys.ens.fr,
      rongvoramnivesvivat@gmail.com,
      sylvain.ribault@ipht.fr,
      paul.roux@phys.ens.fr
    }}
\end{flushleft}
\vspace{7mm}

{\noindent\textsc{Abstract:}
  \vspace{2mm}

  In two-dimensional models of critical non-intersecting loops, there are $\ell$-leg fields that insert $\ell\in\mathbb{N}^*$ open loop segments, and diagonal fields that change the weights of closed loops.
  We conjecture an exact formula for 3-point functions of such fields on the sphere. In the cases of diagonal or spinless 2-leg fields, the conjecture agrees with known results from Conformal Loop Ensembles.

  We numerically compute 3-point functions in loop models on cylindrical lattices, using transfer matrix techniques. The results agree with the conjecture in almost all cases. We attribute the few discrepancies to difficulties that can arise in our lattice computation when the relevant modules of the unoriented Jones-Temperley--Lieb algebra have degenerate ground states.}

\clearpage

\hrule 
\tableofcontents
\vspace{5mm}
\hrule
\vspace{5mm}
\hypersetup{linkcolor=blue!70!black}

\section{Introduction and main results}\label{sec:intro}

Critical loop models were originally introduced as critical limits of \textbf{integrable lattice models}, then studied using the techniques of \textbf{conformal field theory}, and more recently constructed as \textbf{conformal loop ensembles} in a probabilistic approach. Each one of these three approaches has its own strengths, and it is fruitful to compare their results.

In particular, by comparing conformal bootstrap results with ideas from lattice models, it was recently found that a correlation function can be associated to any given combinatorial map \cite{gjnrs23}. To $N$ given punctures on some Riemann surface, we can generally associate several combinatorial maps and therefore several correlation functions. However, in the case of $3$ punctures on the sphere whose numbers of legs sum to an even integer, the combinatorial map is unique, as we now illustrate in a few cases:
\begin{align}
  \label{fig:cmaps}
  \begin{tikzpicture}[baseline=(base), scale = .5]
    \vvv;
    \draw (a) -- (c);
    \draw (b) -- (c);
    \draw (a) to [out = 75, in = -135] (b);
    \draw (a) to [out = 45, in = -105] (b);
    \node at (1.5, -1.5) {3, 3, 2 legs};
    \node at (1.5, -2.5) {(a)};
  \end{tikzpicture}
  \qquad \qquad 
  \begin{tikzpicture}[baseline=(base), scale = .5]
    \vvv;
    \draw (a) -- (c);
    \draw (a) -- (b);
    \draw (a) to [out = -15, in = -90] (3.5, 0) to [out = 90, in = 15] (a);
    \node at (1.5, -1.5) {4, 1, 1 legs};
    \node at (1.5, -2.5) {(b)};
  \end{tikzpicture}
  \qquad \qquad 
  \begin{tikzpicture}[baseline=(base), scale = .5]
    \vvv;
    \draw (a) to [out = 75, in = -135] (b);
    \draw (a) to [out = 45, in = -105] (b);
    \draw (a) to [out = -15, in = -90] (3.5, 0) to [out = 90, in = 15] (a);
    \draw (a) to [out = -30, in = -90] (3.8, 0) to [out = 90, in = 30] (a);
    \node at (1.5, -1.5) {6, 2, 0 legs};
    \node at (1.5, -2.5) {(c)};
  \end{tikzpicture}
\end{align}
In this article, our aim is to determine the corresponding 3-point function. Since the dependence on the punctures' positions is fixed by conformal invariance, this amounts to determining a constant factor, called the 3-point structure constant.

\subsection{Three approaches to critical loop models}

\subsubsection{Integrable lattice models}

The $Q$-state Potts model and the $O(n)$ model are two generalizations of the Ising model. They were originally formulated in terms of local variables with values in $\{1,\dots, Q\}$ and $\mathbb{R}^n$ respectively. The reformulation in terms of loops allows the parameters $Q,n$ to take arbitrary complex values \cite{fk72}, and leads to exact results for critical exponents \cite{nie82}.

To go beyond critical exponents and compute correlation functions, we can use integrability techniques, or transfer matrices built from diagram algebras \cite{nrj23}. However, it is not easy to deduce exact results in the critical limit, for two main reasons:
\begin{itemize}
\item The lattice breaks some of the symmetries of the CFT that appear in the critical limit, in particular the invariance under rotations and dilations.
\item Lattice calculations are limited to finite (and not very large) lattice sizes. We then have to extrapolate to infinite lattices.
\end{itemize}
Despite these difficulties, the lattice approach plays an important role in the study of the CFT, in particular because:
\begin{itemize}
\item It provides a qualitative picture that often survives in the critical limit. Specifically, the lattice interpretation of primary fields of the Conformal Field Theory as either inserting legs or changing the weight of loops naturally leads to the notion of combinatorial maps. These correctly count dimensions of spaces of solutions of crossing symmetry in the conformal bootstrap approach \cite{rib24}.
\item Numerical lattice results provide independent tests of exact formulas that can be derived by other means. In particular, an exact formula for 3-point functions of diagonal fields (also known in the probabilistic literature as full nesting statistics \cite{acsw21}) has been guessed by taking inspiration from Liouville theory, and successfully compared to lattice results \cite{ijs15}. We will now generalize this to 3-point functions of fields with arbitrarily many legs.
\end{itemize}

\subsubsection{Conformal field theory}

Using CFT techniques, it is possible to compute torus partition functions of critical loop models, and to deduce exact formulas for conformal dimensions of primary fields \cite{fsz87}. The resulting primary fields $V_{(r,s)}$ have left and right conformal dimensions $(\Delta,\bar\Delta)=(\Delta_{(r,s)},\Delta_{(r,-s)})$, where
\begin{align}
  \label{eq:spec}
  \Delta_{(r,s)}=\frac14\left(\beta r-\beta^{-1}s\right)^2 -\frac14\left(\beta-\beta^{-1}\right)^2\ .
\end{align}
Here the parameter $\beta$ is related to the central charge $c$ and to the $O(n)$ model's parameter via
\begin{align}
  c= 13 - 6\beta^2-6\beta^{-2} \quad ,\quad n = -2\cos\left(\pi \beta^2\right) \quad \text{with} \quad \Re\beta^2 > 0 \ .
  \label{cnb}
\end{align}
The values of the Kac indices $r,s$ depend on the particular model, subject to the constraints \cite{rib24} 
\begin{align}
  r\in \frac12\mathbb{N}^* \quad , \quad rs\in \mathbb{Z} \ .
\end{align}
The first constraint says that $\ell=2r$ is a positive integer, interpreted as the number of legs of a lattice field. The second constraint is that the conformal spin $rs$ of $V_{(r,s)}$ is integer, again in agreement with the lattice construction.
The notation $V_{(0, s)}$ with $s\in\mathbb{C}$ can also be used for diagonal fields, which assign the weight $w=2\cos(\pi s)$ to loops around them. Let us summarize the properties of the fields $V_{(r,s)}$ and of the corresponding lattice fields:
\begin{align}
  \renewcommand{\arraystretch}{1.3}
  \begin{array}{|c|c|l|}
    \hline
    r & s & \text{Name and properties}
    \\
    \hline\hline
    \frac12\mathbb{N}^* & 0 & \text{Spinless } 2r\text{-leg}
    \\
    \hline
    \frac12\mathbb{N}^* & \frac{1}{r}\mathbb{Z}^* &  2r\text{-leg with spin}
    \\
    \hline
    0 & \mathbb{C} &  \text{Diagonal (therefore spinless)}
    \\
    \hline
  \end{array}
\end{align}
Conformal bootstrap techniques may be used for computing 4-point functions of such fields on the sphere.
To a given 4-point function of the type 
$\left<\prod_{i=1}^4 V_{(r_i,s_i)}\right>$, we associate a linear system of crossing symmetry equations. It was found that the dimension of the space of solutions is the number of combinatorial maps with 4 punctures, such that puncture $i$ has $2r_i$ legs \cite{gjnrs23}. And analytic formulas for solutions could be derived on a case by case basis \cite{nrj23}. 

These results suggest that critical loop models are exactly solvable. However, to actually solve them, we would need not only to determine 3-point functions, but also to deduce $N$-point functions from 3-point functions. This is still an open problem, so we must emphasize that
\begin{quote}
  \textbf{determining 3-point functions}\quad $\centernot\implies$ \quad \textbf{solving critical loop models}.
\end{quote}
Unlike the spins of the Ising model,
the loops of loop models are non-local objects, so it is not clear why correlation functions should obey the axioms of conformal field theory.
Loop models do obey local conformal symmetry, which explains why $N$-point functions are linear combinations of conformal blocks, with coefficients that we call $N$-point structure constants.
However, relating structure constants at different values of $N$ requires the existence of operator product expansions, which does not follow from local conformal symmetry. OPEs would imply that any 4-point structure constant factorizes into two 3-point structure constants.
From numerical \cite{rib22} and analytic \cite{nrj23} results, we can investigate the factorization of 4-point structure constants. In general, 4-point structure constants are finite sums of factorized terms, which suggests that critical loop models do not have OPEs, but some more complicated structures.

\subsubsection{Conformal loop ensembles}

Conformal loop ensembles provide a probabilistic construction of loops in the continuum, without taking a limit from a lattice. 
The construction depends on a parameter $\kappa$, which is related to the CFT parameters by $\kappa = 4\beta^{-2}$, obtained by comparing the Hausdorff dimensions computed from SLE to the spectrum \eqref{eq:spec}.
Conformal loop ensembles were first constructed for the case $\kappa =6$ (percolation) in \cite{cn06} and later for more generic $\kappa$ in \cite{she06}, see also \cite{sw11}. In general, $\kappa$ belongs to the open interval:
\begin{align}
  \frac83< \kappa < 8 \quad \iff \quad \frac12 < \beta^2 < \frac32 \ .
  \label{kappa}
\end{align}
The resulting collections of loops are countably infinite, and become finite in the limits  $\kappa \to\frac83$ (no loops) and $\kappa \to 8$ (a single space-filling loop). In the dilute phase $1<\beta^2 < \frac{3}{2}$, loops are simple curves almost surely (i.e. with probability $1$). In the dense phase $\frac{1}{2} < \beta^2 < 1$, loops almost surely have double points where they touch themselves. The range of $\kappa$ is much smaller than the CFT domain $\Re \beta^2>0$. This is a price to pay for proving theorems, instead of only having physicists' results.

Conformal loop ensembles lead to formulas that agree with lattice models and CFT. For example, the Hausdorff dimension (= the fractal dimension) of a CLE loop \cite{rs04, b04, b08},  is $1+\frac{\kappa}{8}=2-2\Delta_{(1,0)}$, while the dimension of the gasket is $1+\frac{2}{\kappa}+\frac{3\kappa}{32}=2-2\Delta_{(0,\frac12)}$. (For a loop ensemble in a disc, the gasket is the set of points not surrounded by any loops.)
Conformal loop ensembles have recently led to exact results that were not known from the other approaches. This is in particular the case with the backbone exponent \cite{nqsz23}. A recent result that we find particularly stimulating
is the probability that a loop goes through 3 given points \cite[2024 version]{acsw21}. As we will explain in Section~\ref{sec:pi}, this can be interpreted in terms of the 3-point function $\left<V_{(1,0)}V_{(1,0)}V_{(1,0)}\right>$. The formula for $\left<V_{(0,s)}V_{(1,0)}V_{(1,0)}\right>$ is also known \cite[2024 version]{acsw21}.
Our main aim is to generalize these results to all 3-point functions of the type $\left<V_{(r_1,s_1)} V_{(r_2,s_2)} V_{(r_3,s_3)}\right>$.

\subsection{A conjecture inspired by bootstrap results}

In order to derive analytic formulas for 4-point structure constants, it was crucial to have an analytic ansatz to which numerical results could be compared \cite{nrj23}. This ansatz is built from a quantity $C^\text{ref}$ called the reference 3-point structure constant. We now propose to use this quantity as an ansatz for 3-point structure constants. And we rename it $C$ instead of $C^\text{ref}$. Its explicit expression is
\begin{align}
  \boxed{C_{(r_1,s_1)(r_2,s_2)(r_3,s_3)} =\prod_{\epsilon_1,\epsilon_2,\epsilon_3=\pm} \Gamma_\beta^{-1} \left(\tfrac{\beta+\beta^{-1}}{2} + \tfrac{\beta}{2}\left|\textstyle{\sum_i} \epsilon_ir_i\right| + \tfrac{\beta^{-1}}{2}\textstyle{\sum_i} \epsilon_is_i\right)}\ ,
  \label{cref}
\end{align}
where $\Gamma_\beta^{-1}=\frac{1}{\Gamma_\beta}$ is the inverse of the Barnes double Gamma function, which obeys
\begin{align}
  \frac{\Gamma_\beta(x+\beta)}{\Gamma_\beta(x)} = \sqrt{2\pi}\frac{\beta^{\beta x-\frac12}}{\Gamma(\beta x)}
  \quad , \quad 
  \frac{\Gamma_\beta(x+\beta^{-1})}{\Gamma_\beta(x)} = \sqrt{2\pi}\frac{\beta^{\frac12-\beta^{-1}x}}{\Gamma(\beta^{-1}x)} \ .
  \label{gshift}
\end{align}
This special function was not publicly available in free or commercial software, until we implemented it in Julia \cite{roux24}.
The ansatz \eqref{cref} is a simple solution of the following constraints:
\begin{itemize}
\item it obeys shift equations \cite[(3.30)]{rib24} up to a sign,
\item it is symmetric under permutation of the three fields,
\item it is invariant under space parity $(r_i, s_i) \to (-r_i, s_i)$,
\item for 3 diagonal fields $r_i = 0, s_i \in \mathbb{C}$, it reduces to the unique solution of 2 independent shift equations, which coincides with the 3-point structure constant of $c\leq 1$ Liouville theory \cite[(3.38)]{rib24}.
\end{itemize}
In order to interpret and use our ansatz, we need to deal with a few subtleties:
\begin{enumerate}
 \item In addition to $C$, 4-point structure constants involve other factors that are rational fractions in $n$~\eqref{cnb}. In the case of 3-point structure constants, we claim that no such factors are needed.
 \item In order to facilitate the comparison with results from the lattice or from conformal loop ensembles, we normalize our ansatz such that it is invariant under field renormalizations $V_{(r,s)}\to \lambda_{(r,s)}V_{(r,s)}$, including renormalisation of the identity field $V_{(0, 1-\beta^2)}$, which we denote with a subscript $0$:
\begin{align}
  \boxed{\omega_{123} = C_{123} \sqrt{\frac{C_{000}}{C_{011}C_{022}C_{033}}}}\ .
  \label{wott}
\end{align}
The factors in the denominator may be interpreted as 2-point structure constants, explicitly
\begin{align}
  C_{(0,1-\beta^2)(r,s)(r,s)} = \prod_\pm \Gamma_\beta^{-2}\left(\beta^{\pm 1}\right)\prod_{\pm,\pm} \Gamma_\beta^{-1}\left(\beta^{\pm 1}+r\beta \pm s\beta^{-1}\right)\ ,
\end{align}
and $C_{000}$ is the partition function.
\item Nontrivial sign factors would be needed for the ansatz $C$ \eqref{cref} to behave correctly under field permutations, and under shifts $s_i\to s_i+2$ \cite{rib24}. We will ignore this subtlety. Firstly, because in the normalized quantity $\omega$ \eqref{wott}, signs are undetermined since the factor under the square root is in general complex. Secondly, because the sign matters for a ratio of structure constants, but not for an individual structure constant, since we can always change signs in the normalisation of the fields without changing the two-point function. And our focus is on 3-point functions, which involve individual structure constants, unlike 4-point functions.
\end{enumerate}
Our main result is that $\omega_{123}$~\eqref{wott} agrees with the critical limit of 3-point functions in lattice loop models. This is already known to hold in the case $\omega_{(0,s_1)(0,s_2)(0,s_3)}$ \cite{ijs15}, which is itself a generalization of the case $\omega_{(0,\frac12)(0,\frac12)(0,\frac12)}$ that was studied by Delfino and Viti \cite{dv10}.
Moreover, in the cases $\omega_{(1,0)(1,0)(1,0)}$ and $\omega_{(0,s)(1,0)(1,0)}$, our formula agrees with the corresponding quantity in a conformal loop ensemble \cite{acsw21}. We will provide direct numerical evidence for our formula in many examples, detailed in sections \ref{sec:res} and \ref{sec:nonzerospin}. We conjecture that it holds for all allowed values of the parameters,
\begin{align}
  r_i\in\frac12\mathbb{N}\quad , \quad r_is_i\in\mathbb{Z}\quad , \quad r_1+r_2+r_3\in\mathbb{N} \ . 
\end{align}

\subsection{Probabilistic interpretation of spinless fields}\label{sec:pi}

The combinatorial maps with legs are valid probabilistic objects, provided that the punctures are properly regularised.
Consider for simplicity the case of spinless fields ($s_i = 0$).
In the regularisation, the $2r_i$ legs produced by each $V_{(r_i,0)}$ must be taken to emerge not from a point, but from a
suitable small disc. Moreover, each open loop segment in the combinatorial map 
must carry unit weight. However, in the context of CLE it is customary to work only in terms of closed loops, not open loop segments.
We now describe how to relate our results to that context.

\subsubsection{Configurations of closed loops}

For $s_i=0$ and $r_i\in\mathbb{N}^*$, we may trade all open loop segments for a number of closed loops, constrained to pass through 3 punctures. The simplest case is the 3-point function $\left<V_{(1,0)}V_{(1,0)}V_{(1,0)}\right>$, which we may interpret in terms of a single closed loop that passes through the 3 punctures, now drawn as red circles rather than small discs:
\begin{align}
  \begin{tikzpicture}[baseline=(base), scale = .5]
    \vvv;
    \draw (a) -- (c) -- (b) -- (a);
  \end{tikzpicture}
  \qquad 
  \longrightarrow
  \qquad 
  \begin{tikzpicture}[baseline=(base), scale = .45]
    \tever
    \draw[thick] plot [smooth cycle] coordinates {(a) (b) (c)};
  \end{tikzpicture}\quad.
\end{align}
Now, in statistical loop models, closed loops come with the weight $n$~\eqref{cnb}. If we trade the three open loop segments for a single closed loop, we obtain an extra factor $n$ in the corresponding statistical sums. The probability that our closed loop passes through the 3 vertices is therefore $n\left<V_{(1,0)}V_{(1,0)}V_{(1,0)}\right>$. Similar factors of $n$ should complement the 2-point functions $C_{0kk}$ that appear in the normalized 3-point function $\omega_{123}$ \eqref{wott}. We therefore define the normalized probability
\begin{align}
  p^{(1)}_{1,1,1} = n^{-\frac12} \omega_{(1,0)(1,0)(1,0)}\ .
\end{align}
This agrees with the normalized probability that was computed in the Conformal Loop Ensemble approach \cite{acsw21}. 

For more general values of $r_i\in\mathbb{N}^*$, there may exist several different ways to trade $r_1+r_2+r_3$ open loop segments for closed loops.
Assuming that the loop weight $n$ is nonzero, i.e. $\beta^2\notin \mathbb{N}+\frac12$,
the resulting number of closed loops can be any integer $m$ such that
\begin{align}
  1 \leq m \leq \frac12(r_1+r_2+r_3)\ .
\end{align}
For example, in the case $r_1 = r_2 = r_3 = 2$, we obtain the normalized probabilities of 4 inequivalent closed loop configurations, with respectively $m=1,2,2,3$:
\begin{align}
  \begin{tikzpicture}[baseline=(base), scale = .45]
    \tever
    \draw[thick] plot [smooth cycle, tension = .5] coordinates {(-1.9, .1) (-.6, 1)(1.9, -.1) (0, -.9) (-2.1, -.1) (-2.5, 1.5) (0, 3.2) (1, 3) (2.2, .2) (1, 1.2)(0, 2.9) };
  \end{tikzpicture}
  \qquad
  \begin{tikzpicture}[baseline=(base), scale = .45]
    \tever
    \draw[thick] plot [smooth cycle] coordinates {(-1.9, .1) (1.9, .1) (0, 2.8)};
    \draw[thick] plot [smooth cycle, tension = 2.5] coordinates {(-2.1, -.1) (2.1, -.1) (0, 3.2)};
  \end{tikzpicture}
  \qquad
  \begin{tikzpicture}[baseline=(base), scale = .45]
    \tever
    \draw[thick] plot [smooth cycle, tension = 1] coordinates {(-1.9, -.2) (0, .4) (1.9, -.2) (0, -.8)};
    \draw[thick] plot [smooth cycle, tension = 1] coordinates {(-2, .1) (-1.8, 2.6) (0, 3.1) (1.8, 2.6)
      (2, .1)(0, 2.9) };
  \end{tikzpicture}
  \qquad
  \begin{tikzpicture}[baseline=(base), scale = .45]
    \tever
    \draw[thick] plot [smooth cycle, tension = 1] coordinates {(-1.9, -.2) (0, .2) (1.9, -.2) (0, -.9)};
    \draw[thick] plot [smooth cycle, tension = 1] coordinates {(-2.1, .1) (-1.7, 1.9) (-.2, 3) (-.8, 1.5)};
    \draw[thick] plot [smooth cycle, tension = 1] coordinates {(2.1, .1) (1.7, 1.9) (.2, 3) (.8, 1.5)};
  \end{tikzpicture}\quad.
\end{align}
In the first and third examples, a closed loop touches itself at a puncture.
In a dense Conformal Loop Ensemble with $4\leq\kappa <8$ i.e. $\frac12< \beta^2\leq 1$, this is called a non-simple loop. This definition can be extended to the dilute regime $\frac83\leq \kappa\leq 4$, by giving a small but nonzero size to the puncture.

For simplicity we assume that the 2-point function $\left< V_{(r_i,0)}V_{(r_i,0)}\right>$ always involves the maximum number $r_i$ of closed loops. Then the normalized probability that the $m$ closed loops pass through the 3 vertices is 
\begin{align}
  p^{(m)}_{r_1,r_2,r_3} & = n^{m-\frac12(r_1+r_2+r_3)} \omega_{(r_1,0)(r_2,0)(r_3,0)}
  \\
                        & = \frac{n^{m-\frac12(r_1+r_2+r_3)} \prod_\pm \Gamma_\beta^{\frac32}\left(\beta^{\pm 1}\right)\prod_{i=1}^3\Gamma_\beta\left(\beta^{\pm 1}+r_i\beta\right) }{\Gamma_\beta^\frac12\left(2\beta-\beta^{-1}\right)\Gamma_\beta^\frac12\left(2\beta^{-1}-\beta\right)
                          \prod_{\pm,\pm} \Gamma_\beta^{2} \left(\tfrac{\beta+\beta^{-1}}{2} + \tfrac{\beta}{2}\left|r_1\pm r_2\pm r_3 \right| \right)} \ .
\end{align}

\subsubsection{Polymers}

We may interpret the 3-point function $\left<V_{(\frac12, 0)}(z_1)V_{(1,0)}(z_2)V_{(\frac12,0)}(z_3)\right>$ in terms of the probability that an open loop with ends at $z_1$ and $z_3$ passes through $z_2$:
\begin{align}
  \begin{tikzpicture}[baseline=(base), scale = .5]
    \node[fill, circle, minimum size = 2mm, inner sep = 0] at (0, 0) (a) {};
    \node[draw, circle, red!70!black] at (3, 1) (b) {};
    \node[fill, circle, minimum size = 2mm, inner sep = 0] at (6, 0) (c) {};
    \draw[thick] plot [smooth, tension = 1] coordinates {(a) (b) (c)};
    \node[below] at (0, -.2) {$z_1$};
    \node[below] at (3, .7) {$z_2$};
    \node[below] at (6, -.2) {$z_3$};
  \end{tikzpicture}.
\end{align}
This open loop describes a polymer in a soup of polymer loops in the limit $n\to 0$, i.e. either $\beta^2\to\frac12$ (dense polymers, loop-erased random walks, uniform spanning trees) \cite{ds87,lsw04,kw11} or $\beta^2\to\frac32$ (dilute polymers, self-avoiding random walks) \cite{ds87}.

\subsubsection{Three-point functions of spinless fields}

Similarly, any normalized 3-point function of the type $\omega_{(r_1,0)(r_2,0)(r_3,0)}$ with $r_i\in\frac12\mathbb{N}^*$ can be interpreted as the normalized probability that a number of open and closed loops pass through 3 given points, provided we introduce the appropriate power of $n$ as a prefactor. The double Gamma function obeys $\beta,x>0\implies \Gamma_\beta(x)>0$, and has poles for $x\in -\beta \mathbb{N} -\beta^{-1}\mathbb{N}$. This implies
\begin{align}
  \frac12<\beta^2<2 \implies \omega_{(r_1,0)(r_2,0)(r_3,0)} >0\ ,
\end{align}
consistent with a probabilistic interpretation. Positivity would even hold for $0<\beta^2<\infty$, if we removed the field-independent factor $\sqrt{C_{000}}$, which vanishes for $\beta^2\in (\mathbb{N}+2)\cup \frac{1}{\mathbb{N}+2}$ due to factors $\sqrt{\Gamma_\beta^{-1}(2\beta-\beta^{-1})\Gamma_\beta^{-1}(2\beta^{-1}-\beta)}$.
On the other hand, a 3-point function $\omega_{(r_1,s_1)(r_2,s_2)(r_3,s_3)}$ with $s_i\neq 0$ can be negative over some range of values of $\beta$.

\subsubsection*{A few explicit examples}

In order to facilitate comparisons with other analytic or numerical approaches,
let us write the normalized 3-point function~\eqref{wott} in a few examples. To begin with,
\begin{align}
  \omega_{(1,0)(1,0)(1,0)}=  \frac{\sqrt{-\sin(\pi\beta^2)\sin(\pi\beta^{-2})}\Gamma(\frac{1}{\beta^2}-1)\Gamma_\beta(\beta+\beta^{-1})^6\Gamma_\beta(2\beta)^2}
  {2\pi^2 \beta^{1-\beta^{-2}}\Gamma_\beta\left(\tfrac{1}{2\beta}+\beta\right)^{6}\Gamma_\beta\left(\tfrac{1}{2\beta}+2\beta\right)^{2}}
  \ .
\end{align}
In the case of dilute polymers, we find
\begin{subequations}
  \begin{align}
    \omega_{(\frac12,0)(\frac12,0)(1,0)}&\underset{\beta^2=\frac32}{=}
                                          \pi^{-1}2^{-\frac{10}{3}}3^\frac14 \Gamma(\tfrac13)^3 
                                          \approx
                                          0.799071001056270 \ ,
    \\
    \omega_{(1,0)(1,0)(1,0)}&\underset{\beta^2=\frac32}{=}
                              \pi^5 2^6 3^{-\frac34}\Gamma(\tfrac13)^{-9}
                              \approx
                              1.20899262768922 \ .
  \end{align}
\end{subequations}
In the case of critical percolation, we find
\begin{subequations}
  \begin{align}
    \omega_{(1,0)(1,0)(1,0)}&\underset{\beta^2=\frac23}{=}
                              \pi^{-\frac32} 2^{-\frac{107}{12}} 3^2 \Gamma \left(\tfrac{1}{3}\right)^6 \left[
                              \frac{\Gamma_{\sqrt{\frac23}}\left({\sqrt{\frac23}}\right)}{\Gamma_{\sqrt{\frac23}}\left(\sqrt{\frac38}\right)}
                              \right]^8
                              \approx
                              0.952359090621803 \ ,
    \\
    \omega_{(2,0)(2,0)(2,0)}&\underset{\beta^2=\frac23}{=}
                              \pi^{\frac{13}{2}}2^{-\frac{79}{12}}3^{\frac{23}{6}}7^2\Gamma\left(\tfrac13\right)^{-10} \left[
                              \frac{\Gamma_{\sqrt{\frac23}}\left({\sqrt{\frac23}}\right)}{\Gamma_{\sqrt{\frac23}}\left(\frac{5}{2\sqrt{6}}\right)}
                              \right]^8
                              \approx
                              1.779967632825404 \ .
  \end{align}
\end{subequations}
In the case of dense polymers, we find
\begin{align}
  \forall r\in \frac12\mathbb{N}^*\ , \quad \lim_{\beta^2\to \frac{1}{2}}n^{-\frac12}\omega_{(r,0)(r,0)(1,0)} =1\ .
  \label{ustlim}
\end{align}

\subsection{Lattice three-point functions}

\vspace{2mm}

\begin{minipage}{0.6\textwidth}
  \setlength{\parindent}{1.5em}
  Consider a cylindrical square lattice of perimeter $L\in\mathbb{N}$ and length $M\in\mathbb{N}$, in units where the lattice spacing is $1$. We insert 3 fields $V_1,V_2,V_3$ on the bottom circle, in the middle of the cylinder, and on the top circle respectively.
  The critical limit where lattice quantities tend to CFT data will be
  \begin{align}
    1 \ll L \ll M \to \infty\ .
  \end{align}
  In particular, $1\ll L$ allows us to insert arbitrarily many legs, while $L \ll M$ ensures that our cylinder is conformally equivalent to a sphere, minus two discs of negligible size.

  Approaching this limit is doable in the transfer matrix approach. Indeed we compute quantities of the type $\langle \mathrm{out}_3 | T^M \mathcal{O}_2 T^M | \mathrm{in}_1 \rangle$, where the size of the transfer matrix $T$ grows exponentially with $L$. So $L$ cannot be too large, while $M$ is not so contrained: typically $L = O(10), M = O(200)$. The challenge will be to use the transfer matrix for constructing quantities that have a finite limit, such that all non-universal contributions coming from the lattice disappear in the limit.
\end{minipage}
\begin{minipage}{0.37\textwidth}
  \begin{align}
    \hspace{.3cm}
    \label{fig:cylinder}
    \begin{tikzpicture}[baseline = (current bounding box.center), rotate=90]
      \node at (0, 0) {$V_1$} (base);
      \node at (5, -.4) {$V_2$};
      \node at (8, 0) {$V_3$};
      \draw plot [smooth] coordinates {(.6, -.3) (5, -.8) (8.4, .7)};
      \draw plot [smooth] coordinates {(.6, 0) (3, -.2) (5, 0)};
      \draw (.5, -.6) to [out = -20, in = 180] (2, -1);
      \draw[dashed] (2, -1) to [out = 0, in = 180] (4, 1);
      \draw (4, 1) to [out = 0, in = 110] (5, 0);
      \draw[blue] (0, 1) -- (8, 1);
      \draw[blue] (0, -1) -- (8, -1);
      \draw[thick, red] (0, 1) to [out = 180, in = 180] (0, -1);
      \draw[thick, red] (0, -1) to [out = 0, in = 0] (0, 1);
      \draw[thick, red, dashed] (8, 1) to [out = 180, in = 180] (8, -1);
      \draw[thick, red] (8, -1) to [out=0, in=0] (8, 1);
      \fill[thick, red] (5, 0) circle (.1);
      \node at (-.9, 0) {$L$};
      \node[right] at (4, -1) {$2M$};
    \end{tikzpicture}
  \end{align}
\end{minipage}
\vspace{1mm}

On the lattice, correlation functions are statistical sums over configurations of loops, where each configuration comes with a certain weight. This weight is the product of the weights of the closed loops, and of phase factors associated to fields $V_{(r,s)}$ with $s\neq 0$. These phases in principle depend on the angles of the legs at the corresponding vertex \cite{gjnrs23}. We however find it more convenient to fix the angles, and let phases be functions of a cyclic permutation that describes how the legs connect to legs from other vertices. Namely, the phase factor for $V_{(r,s)}$ is described by the representation $e^{i \pi s}$ of the cyclic group $\mathbb{Z}_{2r} = \frac{\mathbb{Z}}{2r\mathbb{Z}}$.

Let $Z_{123}$ be the statistical sum associated to a 3-point function.
In section~\ref{subsec:limit}, we will prove that the ratio
\begin{align}
  \label{eq:3}
  C_{1 2 3}(M, L) = \frac{Z_{123}}{Z_{220}} \sqrt{\frac{Z_{202} Z_{000}}{Z_{101} Z_{303}}}
\end{align}
has a finite limit as $M\to\infty$, which is a simple ratio involving two form factors, see Eq.~\eqref{eq:C123}. In the lattice regularisation the middle operator $V_2$ is
normalised differently from the states $V_1$ and $V_3$ at the cylinder ends, but the latter two are treated symmetrically. In particular, we have $Z_{220} = Z_{022}$,
but $Z_{220} \neq Z_{202}$. To obtain a result that is independent of normalisations, we have therefore replaced the ratio \eqref{wott} by the more general \eqref{eq:3},
itself a special case of \cite[eq.\ 9]{ijs15}. Notice in particular that in \eqref{eq:3}, at the ends and at the middle of the cylinder, each operator appears the same number of
times in the numerator and the denominator, so any normalisation factor drops out of the ratio.

If none of the 3 fields have $s=1$, we conjecture that we recover the 3-point function from CFT by taking the limit $L\to\infty$:
\begin{align}
  \label{eq:2}
  \lim_{L \to \infty} \lim_{M \to \infty} C_{123}(M, L) = \omega_{123}, \qquad s_1, s_2, s_3 \neq 1.
\end{align}
In sections~\ref{sec:res} and ~\ref{sec:nonzerospin}, we check this conjecture by numerically computing the statistical sums $Z_{123}$ in the $O(n)$ and $PSU(n)$ loop models.

In cases involving fields $V_{(r, 1)}$, we observe a more complicated finite-size behaviour. This is because the fields $V_{(r,1)}$ and $V_{(r,-1)}$ are related by parity and therefore belong to the same eigenspace of the transfer matrix. But their second Kac indices differ by $2$, so they belong to the same module of the diagram algebra: that module therefore has a degenerate ground state. We observe that a lattice 3-point function that involves such fields may not have a finite limit, or may tend to $\omega_{123}$, or may tend to a combination of $\omega_{123}$ with its images under parity transformations. Namely, let $\Omega_{123}$ be the set of images of $\omega_{123}$ under independent parity transformations of fields of the type $V_{(r,1)}$; we have $|\Omega_{123}|\leq 4$ because $\omega_{123}$ is invariant under the simultaneous transformation of the 3 fields. When the large-size limit exists, we find that it is of the type
\begin{align}
  \label{eq:21}
  \lim_{L \to \infty} \lim_{M \to \infty} \left( \frac{\pi}{L} \right)^{\alpha_{123}} f_{123} C_{123}(M, L) = \sum_{\omega\in\Omega_{123}} \epsilon_\omega \omega \quad \text{with} \quad \left\{\begin{array}{l}\epsilon_\omega\in\{-1,0,1\} \ , \\
  \alpha_{123} \in \{-1, 0, 1\} \ , \\
  f_{123} \in \{ 1, \sqrt2, \frac{1}{\sqrt2} \} \ .
  \end{array} \right.
\end{align}
We provide numerical results in Table \eqref{tab}.

\section{Loop models on cylindrical lattices}\label{sec:cyl}

\begin{minipage}{0.62\textwidth}
  \setlength{\parindent}{1.5em}
  Consider a square lattice ${\mathcal S}_{L,M}$ wrapped on a cylinder so that
  the cylinder axis is parallel to one of the principal directions of the lattice, with $L$ lattice spacings in the periodic direction, and define the discrete spatial coordinate $x=1,2,\ldots,L$ along that direction. In the other, orthogonal direction, along the cylinder, the length is $2M$ lattice spacings, corresponding to a discrete coordinate $t=-M,-M+1,\ldots,M$ of imaginary time. Integer $(x,t)$ correspond to mid-points of vertical edges of ${\mathcal S}_{L, M}$.
\end{minipage}
\begin{minipage}{0.36\textwidth}
  \begin{align}
    \hspace{-.5cm}
    \label{fig:cyl-lat}
    \begin{tikzpicture}[scale=0.5, baseline=(current bounding box.center),
      oriented/.style={postaction={decorate},decoration={markings,
          mark=at position #1 with {\arrow{>}}}}
      ]
      \foreach \i in {1, 2, 3, 4, 5, 6, 7, 8, 9, 10}
      \draw (2, \i-0.5) arc[start angle=0, end angle=180, x radius=2cm, y radius=-0.2cm]; 
      \pgfmathsetmacro{\L}{6}
      \foreach \i in {0, 1, 2, 3, 4, 5, 6} {
        \pgfmathsetmacro{\t}{(6-\i) * 180 / \L}
        \pgfmathsetmacro{\x}{2 * cos(\t)}
        \pgfmathsetmacro{\y}{-0.2 * sin(\t)}
        \draw (\x, \y) -- ++ (0, 10);
      }
      \draw[red, postaction={decorate}, decoration={markings, mark=at position 0.75 with {\arrow[scale=1.3]{Stealth[reversed]}}}]
      (0, 10) ellipse[x radius=2cm, y radius=0.2cm];
      \node[right, red] at (2, 10) {$R_3$};
      \draw[red, red, postaction={decorate}, decoration={markings, mark=at position 0.5 with {\arrow[scale=1.3]{Stealth[reversed]}}}] (2, 0) arc[start angle=0, end angle=180, x radius=2cm, y radius=-0.2cm];
      \node[right, red] at (2, 0) {$R_1$}; 
      \node[left] at (-2, 0) {$_{t=-M}$}; 
      \draw[gray, dashed] (2, 5) arc[start angle=0, end angle=180, x radius=2cm, y radius=-0.2cm]; 
      \node[left] at (-2, 5) {$_{t=0}$}; 
      \node[left] at (-2, 10) {$_{t=M}$}; 
      \draw[red, dashed] (2, 0) arc[start angle=0, end angle=180, x radius=2cm, y radius=0.2cm]; 
      \draw[red, red, postaction={decorate}, decoration={markings, mark=at position 0.75 with {\arrow[scale=1.3]{Stealth[reversed]}}}] (0, 4.8) ellipse[x radius=1.2cm, y radius=5pt];
      \node[red, right] at (2, 4.8) {$R_2$};
      \node[below=3pt] (0, -0.2) {$L$};
      \draw[red,->,thick,>=Stealth] (1,2) arc[start angle=0, end angle=300, radius=0.8cm];
    \end{tikzpicture}
  \end{align}
\end{minipage}

\subsection{The models}\label{subsec:models}

Let us define the $O(n)$ and $PSU(n)$ lattice models on ${\mathcal S}_{L,M}$ \cite{wns92}. At each vertex, we allow 9 configurations, to which we give local vertex weights $\rho_i$:
\begin{align}
  \begin{tikzpicture}[baseline={([yshift=-3pt]current bounding box.center)}, scale = .4]
    \draw[gray] (0,-1) -- (0,1);
    \draw[gray] (-1,0) -- (1,0);
    \node[below] at (0,-1) {$\rho_1$};
  \end{tikzpicture} \qquad
  \begin{tikzpicture}[baseline={([yshift=-3pt]current bounding box.center)}, scale = .4]
    \draw[gray] (0,-1) -- (0,1);
    \draw[gray] (-1,0) -- (1,0);
    \draw[very thick,blue] (-1,0) to (-0.3,0) arc (-90:0:0.3) to (0,1);
    \node[below] at (0,-1) {$\rho_2$};
  \end{tikzpicture} \qquad
  \begin{tikzpicture}[baseline={([yshift=-3pt]current bounding box.center)}, scale = .4]
    \draw[gray] (0,-1) -- (0,1);
    \draw[gray] (-1,0) -- (1,0);
    \draw[very thick,blue] (0,-1) to (0,-0.3) arc (180:90:0.3) to (1,0);
    \node[below] at (0,-1) {$\rho_3$};
  \end{tikzpicture} \qquad
  \begin{tikzpicture}[baseline={([yshift=-3pt]current bounding box.center)}, scale = .4]
    \draw[gray] (0,-1) -- (0,1);
    \draw[gray] (-1,0) -- (1,0);
    \draw[very thick,blue] (0,-1) to (0,-0.3) arc (0:90:0.3) to (-1,0);
    \node[below] at (0,-1) {$\rho_4$};
  \end{tikzpicture} \qquad
  \begin{tikzpicture}[baseline={([yshift=-3pt]current bounding box.center)}, scale = .4]
    \draw[gray] (0,-1) -- (0,1);
    \draw[gray] (-1,0) -- (1,0);
    \draw[very thick,blue] (0,1) to (0,0.3) arc (180:270:0.3) to (1,0);
    \node[below] at (0,-1) {$\rho_5$};
  \end{tikzpicture} \qquad
  \begin{tikzpicture}[baseline={([yshift=-3pt]current bounding box.center)}, scale = .4]
    \draw[gray] (0,-1) -- (0,1);
    \draw[gray] (-1,0) -- (1,0);
    \draw[very thick,blue] (-1,0) to (1,0);
    \node[below] at (0,-1) {$\rho_6$};
  \end{tikzpicture} \qquad
  \begin{tikzpicture}[baseline={([yshift=-3pt]current bounding box.center)}, scale = .4]
    \draw[gray] (0,-1) -- (0,1);
    \draw[gray] (-1,0) -- (1,0);
    \draw[very thick,blue] (0,-1) to (0,1);
    \node[below] at (0,-1) {$\rho_7$};
  \end{tikzpicture} \qquad
  \begin{tikzpicture}[baseline={([yshift=-3pt]current bounding box.center)}, scale = .4]
    \draw[gray] (0,-1) -- (0,1);
    \draw[gray] (-1,0) -- (1,0);
    \draw[very thick,blue] (-1,0) to (-0.3,0) arc (-90:0:0.3) to (0,1);
    \draw[very thick,blue] (0,-1) to (0,-0.3) arc (180:90:0.3) to (1,0);
    \node[below] at (0,-1) {$\rho_8$};
  \end{tikzpicture} \qquad
  \begin{tikzpicture}[baseline={([yshift=-3pt]current bounding box.center)}, scale = .4]
    \draw[gray] (0,-1) -- (0,1);
    \draw[gray] (-1,0) -- (1,0);
    \draw[very thick,blue] (0,-1) to (0,-0.3) arc (0:90:0.3) to (-1,0);
    \draw[very thick,blue] (0,1) to (0,0.3) arc (180:270:0.3) to (1,0);
    \node[below] at (0,-1) {$\rho_9$};
  \end{tikzpicture}
\end{align}
In addition there is  a non-local weight $n$ for each closed loop formed by contingent pieces of blue curves. 
The partition function reads
\begin{align}
  Z = \sum_{\mathrm{configurations}} n^{\# {\mathrm{loops}}} \prod_{i=1}^9 \rho_i^{V_i} \,,
\end{align}
where $V_i$ is the number of vertices of ${\mathcal S}_{L,M}$ having the local configuration $i$.

The $O(n)$ model is quantum integrable for a specific choice of the
weights, depending on a crossing parameter $\lambda = \frac{\pi \beta^2}{4}$ and a spectral parameter $u$. There exists a choice of the spectral parameter such that these weights acquire the discrete rotational invariance of the square lattice:
\begin{align}
  \label{eq:9}
  \rho_2 = \rho_3 = \rho_4 = \rho_5, \quad \rho_6 = \rho_7, \quad \rho_8 = \rho_9.
\end{align}
For this specific choice, the weights are
\begin{subequations}
  \begin{align}
    \rho_1 &= 1 + \sin\left(\tfrac{\pi\beta^2}{4}\right) + \sin\left(\tfrac{3\pi\beta^2}{4}\right) - \sin\left( \tfrac{5\pi\beta^2}{4}\right) \,, \\
    \rho_2 &= \rho_3 = \rho_4 = \rho_5 = 2 \sin\left(\tfrac{\pi\beta^2}{2}\right) \sin \left( \tfrac{\pi}{4}\left( \tfrac{3}{2}\beta^2+1 \right) \right) \,, \\
    \rho_6 &= \rho_7 = 1 + \sin\left(\tfrac{3\pi\beta^2}{4}\right) \,, \\
    \rho_8 &= \rho_9 = \sin\left(\tfrac{\pi\beta^2}{4}\right) + \cos\left(\tfrac{\pi\beta^2}{2}\right) \,.
  \end{align}
\end{subequations}
The continuum limit has been shown to be conformally invariant for real $\beta$, and by periodicity and symmetry of the weights we can
focus on the interval $\beta^2 \in [0,2]$. In particular, $\beta^2 \in [0,1]$ defines the \textbf{dense phase} of the $O(n)$ model with a finite density of monomers, while $\beta^2 \in [1, 2]$ defines the \textbf{dilute phase}.
We assume $\beta^2\in[\frac12,\frac32]$ as in Eq.~\eqref{kappa}, then $n,\rho_i\geq 0$.

The $PSU(n)$ model is defined by forbidding 7 of the 9 local configurations, such that the loops are completely packed:
\begin{align}
  \rho_1 = \cdots = \rho_7 = 0 \quad ,\quad
  \rho_8 = \rho_9 = 1 \ .
\end{align}
These weights do not depend on $\beta^2$, so the model is invariant under $\beta^2\to \beta^2+1$, just like $n$~\eqref{cnb}. We therefore restrict to $\beta^2 \in [0,1]$.
This model has a $PSU(n)$ symmetry group, which is larger than $O(n)$ \cite{rjrs24}. It is closely
related to the $Q = n^2$ state Potts model in its formulation as a Fortuin-Kasteleyn cluster model \cite{fk72}.
Indeed, defining the Potts spins on a rotated square lattice of which ${\mathcal S}_{L,M}$ is the medial (or surrounding) lattice \cite{bkw76}, the loops are the domain walls between clusters and dual clusters.

\subsection{Transfer matrix}\label{subsec:trm}

Let us build a transfer matrix $T$ that propagates between a row at time $t$ and an adjacent row at time $t+1$.
The evolution up to time $t$ is encoded in a link pattern $|\sigma_t\rangle$: a set of non-intersecting arcs indicating which of the $L$ points at row $t$ are connected together in the part of the cylinder having time coordinates in $[-M,t]$:
\begin{subequations}
  \begin{align}
    \label{fig:sqlat}
    &\begin{tikzpicture}[scale=0.9, baseline=(current bounding box.center)]
      \draw[-stealth] (-0.5, -0.5) -- (4.5, -0.5) node[right]{$x$};
      \draw[-stealth] (-0.5, -0.5) -- (-0.5, 4.5) node[left]{$t$};
      \draw[black!40] (-0.5, -0.5) grid (4.3, 4.3);
      \draw[fill] (-0.5, 0+0.5) node[left]{$_{t_0}$} circle (1pt);
      \draw[fill] (0, -0.5) node[below]{$_0$} circle (1pt);
      \foreach \i in {1, 2, 3}
      {
        \draw[fill] (-0.5, \i+0.5) node[left]{$_{t_0+\i}$} circle (1pt);
        \draw[fill] (\i, -0.5) node[below]{$_{\i}$} circle (1pt);
      }
      \draw[fill] (4, -0.5) node[below]{$_4$} circle (2pt);
      \draw[loops] (0, 0) -- (0, 1) -- (1, 1) -- (1, 0) -- cycle;
      \draw[loops] (2, 1) -- (3, 1) -- (3, 3) -- (2, 3) -- cycle;
      \draw[loops] (1, 1) -- (1, 4) -- (4, 4) -- (4, 1) -- (3, 1) -- (3, 0) -- (2, 0) -- (2, 1) -- cycle;
      \draw[red, dashed] (-0.5, 2.5) -- (4.5, 2.5) node (t2){};
      \draw[red, dashed] (-0.5, 3.5) -- (4.5, 3.5) node (t3){};
      \begin{scope}[xshift=8cm, yshift=2cm]
        \node[anchor = north east] (lp2) at (0, 0) {$|\sigma_{t_0+2}\rangle = $};
        \foreach \i in {0, 1, 2, 3, 4}
        \draw[fill] (\i, 0) circle (2pt);
        \draw[thick, blue] (1, 0) to[out=-90, in=-90] (4, 0);
        \draw[thick, blue] (2, 0) to[out=-50, in=-130] (3, 0);
      \end{scope}
      \begin{scope}[xshift=8cm, yshift=4cm]
        \node[anchor = north east, fill=none] (lp3) at (0, 0) {$|\sigma_{t_0+3}\rangle = $};
        \foreach \i in {0, 1, 2, 3, 4}
        \draw[fill] (\i, 0) circle (2pt);
        \draw[thick, blue] (1, 0) to[out=-90, in=-90] (4, 0);
      \end{scope}
      \draw[-stealth, red] (t2) -- (lp2);
      \draw[-stealth, red] (t3) -- (lp3);
    \end{tikzpicture}
  \end{align}
  While this example is given for a fixed configuration on ${\mathcal S}_{L,M}$, the goal is to compute the sum over all configurations with their proper statistical weights.
  We therefore define the state at time $t$ to be a linear combination of all possible $|\sigma_t\rangle$, with a coefficient equal to the product of the local weights
  of vertices in the rows $[-M,t]$  times the weights of loops having been completed prior to $t$. Note that this depends on $|\sigma_t\rangle$ and on
  the boundary condition at time $-M$.
  The transfer matrix $T$ is the linear combination of all possible vertex configurations at time $t+\tfrac12$ with weights $\prod_V\rho_V$, seen as a linear operator that produces the state at time $t+1$ from that at $t$. In particular, the term that produces $|\sigma_{t_0+3}\rangle$ from $|\sigma_{t_0+2}\rangle$ is
  \begin{align}
    & T = \rho_1\rho_3\rho_4\rho_7^2 \quad
      \begin{tikzpicture}[baseline=(current bounding box.center)]
        \draw[black!40] (-0.5, -0.4) grid (4.5, 0.4);
        \draw[loops] (1, -0.5) -- (1, 0.5);
        \draw[loops] (4, -0.5) -- (4, 0.5);
        \draw[loops] (2, -0.5) -- (2, 0) -- (3, 0) -- (3, -0.5);
        \draw[red, dashed] (-0.5, -0.5) -- (4.5, -0.5);
        \draw[red, dashed] (-0.5, 0.5) -- (4.5, 0.5);
      \end{tikzpicture}
      \quad  + \cdots
    \\ & \implies
         T |\sigma_{t_0+2}\rangle = n \rho_1\rho_3\rho_4\rho_7^2 |\sigma_{t_0+3}\rangle + \cdots \ ,
         \label{eq:Tsigma}
  \end{align}
\end{subequations}
where the factor of $n$ is here because the action of $T$ closes a loop.

The transfer matrix $T$ belongs to the \textbf{unoriented Jones-Temperley-Lieb algebra} $\uJTL_L(n)$ \cite{rjrs24} in the case of the $PSU(n)$ model, and a dilute version thereof in the case of the $O(n)$ model. The algebra acts by diagram concatenation on link patterns, which span a module $\mathcal{W}_0$ of that algebra. In the $PSU(n)$ model, link patterns cannot have empty sites, which implies $L \in 2 \mathbb{N}$.

Given two link patterns $|\psi_1\rangle,|\psi_2\rangle$, let $\langle \psi_1|\otimes |\psi_2\rangle$ be the diagram obtained by concatenating $|\psi_2\rangle$ with the time reflection of $|\psi_1\rangle$. We define a bilinear form $\langle \cdot | \cdot \rangle$ on $\mathcal{W}_0$ by
\begin{align}
  \langle \psi_2\mid\psi_1 \rangle = 0^{\#\text{open loops in }\langle \psi_1|\otimes |\psi_2\rangle} n^{\#\text{closed loops in }\langle \psi_1|\otimes |\psi_2\rangle}\ .
  \label{eq:psi0}
\end{align}
For instance,
\begin{subequations}
  \begin{align}
    \label{eq:12}
    &\left\langle\hspace{0.3cm}
      \begin{tikzpicture}[scale=0.5, baseline=(base)]
        \coordinate (base) at (0, -.5);
        \foreach \i in {0, 1, 2, 3, 4}
        \draw[fill] (\i, 0) circle (2pt);
        \draw[thick, blue] (1, 0) to[out=-90, in=-90] (4, 0);
        \draw[thick, blue] (2, 0) to[out=-50, in=-130] (3, 0);
      \end{tikzpicture}\hspace{0.3cm}
      \middle| \hspace{0.3cm}
      \begin{tikzpicture}[scale=0.5, baseline=(base)]
        \coordinate (base) at (0, -.5);
        \foreach \i in {0, 1, 2, 3, 4}
        \draw[fill] (\i, 0) circle (2pt);
        \draw[thick, blue] (1, 0) to[out=-90, in=-90] (4, 0);
      \end{tikzpicture}\hspace{0.3cm}
      \right\rangle =
      \begin{tikzpicture}[scale=0.5, baseline=(current bounding box.center)]
        \foreach \i in {0, 1, 2, 3, 4}
        \draw[fill] (\i, 0) circle (2pt);
        \draw[thick, blue] (1, 0) to[out=90, in=90] (4, 0);
        \draw[thick, blue] (1, 0) to[out=-90, in=-90] (4, 0);
        \draw[thick, blue] (2, 0) to[out=-50, in=-130] (3, 0);
      \end{tikzpicture}
      = 0\ , \\
    &\left\langle\hspace{0.3cm}
      \begin{tikzpicture}[scale=0.5, baseline=(base)]
        \coordinate (base) at (0, -.5);
        \foreach \i in {0, 1, 2, 3, 4}
        \draw[fill] (\i, 0) circle (2pt);
        \draw[thick, blue] (1, 0) to[out=-90, in=-90] (2, 0);
        \draw[thick, blue] (3, 0) to[out=-90, in=-90] (4, 0);
      \end{tikzpicture} \hspace{0.3cm}
      \middle| \hspace{0.3cm}
      \begin{tikzpicture}[scale=0.5, baseline=(base)]
        \coordinate (base) at (0, -.5);
        \foreach \i in {0, 1, 2, 3, 4}
        \draw[fill] (\i, 0) circle (2pt);
        \draw[thick, blue] (1, 0) to[out=-90, in=-90] (4, 0);
        \draw[thick, blue] (2, 0) to[out=-50, in=-130] (3, 0);
      \end{tikzpicture} \hspace{0.3cm}
      \right\rangle =
      \begin{tikzpicture}[scale=0.5, baseline=(current bounding box.center)]
        \foreach \i in {0, 1, 2, 3, 4}
        \draw[fill] (\i, 0) circle (2pt);
        \draw[thick, blue] (1, 0) to[out=90, in=90] (2, 0);
        \draw[thick, blue] (3, 0) to[out=90, in=90] (4, 0);
        \draw[thick, blue] (1, 0) to[out=-90, in=-90] (4, 0);
        \draw[thick, blue] (2, 0) to[out=-50, in=-130] (3, 0);
      \end{tikzpicture}
      = n\ .
  \end{align}
\end{subequations}
For this bilinear form, the transfer matrix is invariant with respect to $\left\langle \cdot | \cdot \right\rangle$, i.e. $\langle Tv | w \rangle = \langle v | Tw \rangle$ or equivalently $T^\dagger = T$, where $^\dagger$ is linear and acts on diagrams and on link patterns by reflecting them with respect to the horizontal axis.
We could have chosen $\langle \cdot | \cdot \rangle$ to be a sesquilinear form instead, in which case $^{\dagger}$ would have been antilinear. We find the choice of a bilinear form more convenient, especially when $n \in \mathbb{C}$.
In particular, the following two definitions of the partition function are equivalent,
\begin{align}
  \label{eq:Z000}
  Z = \langle \psi_0 | T^{2M} \psi_0 \rangle = \langle T^M \psi_0 | T^M \psi_0 \rangle \,,
\end{align}
where we use the following vacuum states, which serve as boundary conditions at the top and bottom of the cylinder:
\begin{align}
  \label{eq:13}
  \left|\psi_0^{O(n)}\right\rangle  \hspace{1cm}&= \hspace{1cm}
                                         \begin{tikzpicture}[scale=0.5]
                                           \foreach \i in {0, 1, 2, 3, 6, 7}
                                           \draw[fill] (\i, 0) circle (2pt);
                                           \node at (4.5, 0) {$\dots$};
                                         \end{tikzpicture}
  \\
  \left|\psi_0^{PSU(n)}\right\rangle \hspace{1cm}&= \hspace{1cm}
                                          \begin{tikzpicture}[scale=0.5]
                                            \foreach \i in {0, 1, 2, 3, 6, 7}
                                            \draw[fill] (\i, 0) circle (2pt);
                                            \draw[thick, blue] (0, 0) [out=-90, in= -90] to (1, 0);
                                            \draw[thick, blue] (2, 0) [out=-90, in= -90] to (3, 0);
                                            \draw[thick, blue] (6, 0) [out=-90, in= -90] to (7, 0);
                                            \node at (4.5, 0) {$\dots$};
                                          \end{tikzpicture}
\end{align}
An element of $\mathcal{W}_0$ such as $T^M|\psi_0\rangle$ is in general a linear combination of $d = \dim \mathcal{W}_0$ link patterns. Therefore, evaluating the second bilinear form in~\eqref{eq:Z000} requires $d^2$ operations, while the first one requires only $d$ operations and is numerically vastly more efficient.

\subsection{Three-point functions of spinless fields}
\label{subsec:3ptspinless}

In order to describe the three-point function $\langle V_1V_2V_3\rangle$, we will use lattice quantities of the type
\begin{align}
  Z_{123} = \langle \psi_3 | T^M \mathcal{O}_2 T^M | \psi_1 \rangle \ .
  \label{eq:17}
\end{align}
Here the boundary states $|\psi_1\rangle,|\psi_3\rangle$ are inserted at the bottom and top of the cylinder, while the operator $\mathcal{O}_2$ is in the middle. In the critical limit, these states and operators are supposed to describe the conformal fields $V_i$. By universality, this leaves us with a lot of possibilities for constructing these objects. Our choices of constructions will be guided by the representation theory of the $\uJTL_L(n)$ algebra, and by considerations of numerical efficiency.

We will describe our constructions of boundary states and operators in several cases of increasing technical difficulty. The basic idea is that a state or operator that describes a field $V_{(r,s)}$ has $\ell=2r$ legs, as in the combinatorial maps~\eqref{fig:cmaps}.

The corresponding modules of the $\uJTL_L(n)$ algebra are indexed by two numbers $(r, s)$, such that $rs \in \mathbb{Z}$, $s \in (-1, 1]$. The case $r=0$ is the module $\mathcal{W}_0$ of section \ref{subsec:trm}, in which case $s$ is not defined. The modules $\mathcal{W}_{(r>0, s)}$ are spanned by link patterns with $2r$ defect lines, as defined in \eqref{eq:lp_def}, and the action of the algebra is such that the $2r$ defect lines carry a representation $e^{i \pi s}$ of the cyclic group $\mathbb{Z}_{2r}$ \cite{gl98}, \cite{js18}.

In order to compute the three-point functions of \eqref{eq:17}, we need to keep track of the origin of the legs, in order in particular to allow for some of the defects coming from the $| \psi_1 \rangle$ insertion to connect with those inserted by $\mathcal{O}_2$. To do this, we have to add extra information to link patterns: as little information as necessary, if we care about numerical efficiency.
The simplest case is the case of spinless leg fields that have no \textbf{enclosures}, i.e.\ no legs that connect one and the same field.
For any permutation $\{i, j, k\}$ of $\{1, 2, 3\}$, the field $i$ has enclosures if $\ell_i > \ell_j+\ell_k$: the number of enclosures is then $\frac12(\ell_i-\ell_j-\ell_k)=r_i-r_j-r_k$.

The modules spanned by link patterns with labelled defects are reducible. The operator $\mathcal{O}_2$ defined in this section, and its spin version $\mathcal{O}_{(r_2, s_2)}$ defined in Section~\ref{subsec:3ptspin} are morphisms
\begin{align}
\label{eq:26}
\mathcal{W}_{(r_1, s_1)} \to \left\{ \text{labelled link patterns with }\ell \in \{|\ell_1 - \ell_2|,\dots,|\ell_1 + \ell_2| \}\right\}.
\end{align}
Since the operators $\mathcal{O}$ do not commute with the action of the algebra, we cannot redefine them as morphisms between irreducible modules. 
\\
The case of nonzero spins $rs\neq 0$ is postponed to 

\subsubsection{$\langle V_{(r_1,0)} V_{(r_2,0)} V_{(r_3,0)} \rangle$ with no enclosure}

The module $\mathcal{W}_{(r, 0)}$ of the $\uJTL_L(n)$ algebra is
\begin{align}
  \label{eq:19}
  \mathcal{W}_{(r, 0)} = \operatorname{Span}(\{ \text{link patterns with } \ell = 2r \text{ defects} \}),
\end{align}
where link patterns with defects are represented as
\begin{align}
  \label{eq:lp_def}
  \begin{tikzpicture}[scale=0.8]
    \draw[black!40] (-0.5, 0) grid (5.3, 4.3);
    \draw[loops] (0, 0) -- (0, 3) -- (1, 3) -- (1, 4.3);
    \draw[loops] (1, 0) -- (1, 3) -- (2, 3) -- (2, 1) -- (3, 1) -- (3, 0) -- (4, 0) -- (4, 1) -- (5, 1) -- (5, 4.3);
    \draw[loops] (3, 1) -- (3, 3) -- (4, 3) -- (4, 1) -- cycle;
    \draw[red, dashed] (-0.5, 2.5) node[black, left] {$t$} -- (5.5, 2.5) node (t2){};
    \draw[red, dashed] (-0.5, 3.5) node[black, left] {$t+1$} -- (5.5, 3.5) node (t3){};
    \begin{scope}[xshift=9cm, yshift=1cm]
      \node[anchor = north east] (lp2) at (0, 0) {$\sigma_t = $};
      \foreach \i in {0, 1, 2, 3, 4, 5}
      \draw[fill] (\i, 0) circle (2pt);
      \draw[thick, blue] (0, 0) -- ++(0, -1); 
      \draw[thick, blue] (1, 0) -- ++(0, -1); 
      \draw[thick, blue] (2, 0) to[out=-90, in=-90] (5, 0);
      \draw[thick, blue] (3, 0) to[out=-50, in=-130] (4, 0);
    \end{scope}
    \begin{scope}[xshift=10cm, yshift=4cm]
      \node[anchor = north east, fill=none] (lp3) at (0, 0) {$\sigma_{t+1} = $};
      \foreach \i in {0, 1, 2, 3, 4, 5}
      \draw[fill] (\i, 0) circle (2pt);
      \draw[thick, blue] (1, 0) -- ++(0, -2);
      \draw[thick, blue] (5, 0) -- ++(0, -2);
    \end{scope}
    \draw[-stealth, red] (t2) -- (lp2);
    \draw[-stealth, red] (t3) -- (lp3);
  \end{tikzpicture}
\end{align}
and the action of the transfer matrix is defined by diagram concatenation. In addition, each defect carries a label $1$ or $2$, depending on whether its origin is in the state $|\psi_1\rangle$ or in the operator $\mathcal{O}_2$. The boundary condition at the bottom of the cylinder is now
\begin{subequations}
  \label{eq:14}
  \begin{align}
    | \psi_1^{O(n)} \rangle &=
                     \begin{tikzpicture}[scale=0.7, baseline=(current bounding box.center)]
                       \foreach \i in {0, 1, 3, 4, 5}
                       \draw[fill] (\i, 0) circle (2pt);
                       \node at (2, 0) {$\dots$};
                       \foreach \i in {0, 1, 3}
                       \draw[thick, blue] (\i, 0) -- ++(0, -1) node[below, black]{$_1$};
                       \node at (6, 0) {$\dots$};
                       \draw [decorate,decoration={brace,amplitude=5pt}] (0, 5pt) -- node[midway, above, anchor=south]{$\ell_1$ defects}(3, 5pt);
                     \end{tikzpicture}
    \\
    |\psi_1^{PSU(n)} \rangle &=
                      \begin{tikzpicture}[scale=0.7, baseline=(current bounding box.center)]
                        \foreach \i in {0, 1, 3, 4, 5, 6, 7}
                        \draw[fill] (\i, 0) circle (2pt);
                        \node at (2, 0) {$\dots$};
                        \foreach \i in {0, 1, 3}
                        \draw[thick, blue] (\i, 0) -- ++(0, -1) node[below, black]{$_1$};
                        \node at (8, 0) {$\dots$};
                        \draw[thick, blue] (4, 0) to[out=-90, in=-90] (5, 0);
                        \draw[thick, blue] (6, 0) to[out=-90, in=-90] (7, 0);
                      \end{tikzpicture}.
  \end{align}
\end{subequations}
An operator $\mathcal{O}_2$ with $\ell_2=2r_2$ legs is represented as
\begin{align}
  \label{eq:15}
  \mathcal{O}_2 &=
        \begin{tikzpicture}[scale=0.7, baseline=(current bounding box.center)]
          \draw[red, dashed] (-0.3, 0) -- (6.3, 0);
          \draw[red, dashed] (-0.3, 2) -- (6.3, 2);
          \foreach \i in {0, 1, 3} {
            \draw[fill] (\i, 0.9) circle (2pt);
            \draw[thick, blue] (\i, 0.9) -- ++(0, -0.9);
            \draw[fill] (\i, 1.1) circle (2pt);
            \draw[thick, blue] (\i, 1.1) -- ++(0, 0.9);
            \node[below] at (\i, 0) {$_2$};
            \node[above] at (\i, 2) {$_2$};
          }
          \draw[red, rounded corners=4pt] (-0.2, 0.8) -- (3.2, 0.8) -- (3.2, 1.2) -- (-0.2, 1.2) -- cycle;
          \node at (2, 1) {$\dots$};
          \foreach \i in {4, 6}
          \draw[thick, blue, dashed] (\i, 0) -- (\i, 2);
          \node at (5, 1) {$\dots$};
        \end{tikzpicture} \quad ,
\end{align}
where the edges $x=1,2,\ldots,\lfloor \ell_2/2 \rfloor$ are broken into two half-edges, one pointing towards the bottom and the other pointing towards the top. In the $PSU(n)$ model, $\ell_i\in 2\mathbb{N}$ must be even. 
In the $O(n)$ model, $\ell_i\in\mathbb{N}$ may be any integer. If $\ell_2$ is odd, the edge $x=(\ell_2+1)/2$ is also broken, and we take one of the resulting half-edges to be occupied and the other to be empty. In order to have a time-reversal symmetric situation, we symmetrise over the cases where this last occupied half-edge points towards the bottom and the top, respectively.

The action of $\mathcal{O}_2$ on a link pattern is given by diagram concatenation, with the rule that $2$-defects are allowed to be contracted with $1$-defects. But defects with identical labels  cannot be contracted among themselves, which happens when $\mathcal{O}_2$ acts on a link pattern which has an arch between positions $\leq \lfloor \frac{\ell_2}{2} \rfloor$: in this case we set the result to 0.
On each of the remaining edges, $x=\lfloor \ell_2/2 \rfloor+1,\ldots,L$, the operator ${\mathcal O}_2$ acts as the identity operator, and so can connect two occupied or two empty edges. We represent this as dashed blue lines in the diagram.
For instance, in the case $\ell_2=4$,
\begin{subequations}
  \begin{align}
    \label{eq:16}
    &\mathcal{O}_2 \cdot
      \begin{tikzpicture}[scale=0.5, baseline=(current bounding box.center)]
        \foreach \i in {0, 1, 2, 3}
        \draw[fill] (\i, 0) circle (2pt);
        \draw[thick, blue] (0, 0) to[out=-90, in=-90] (1, 0);
        \draw[thick, blue] (2, 0) to[out=-90, in=-90] (3, 0);
      \end{tikzpicture}
      =
      \begin{tikzpicture}[scale=0.5, baseline=(current bounding box.center)]
        \foreach \i in {0, 1, 2, 3}
        \draw[fill] (\i, 0) circle (2pt);
        \draw[thick, blue] (0, 0) to[out=-90, in=-90] (1, 0);
        \draw[thick, blue] (2, 0) to[out=-90, in=-90] (3, 0);
        \foreach \i in {0, 1} {
          \draw[fill] (\i, 0.9) circle (2pt);
          \draw[thick, blue] (\i, 0.9) -- ++(0, -0.9);
          \draw[fill] (\i, 1.1) circle (2pt);
          \draw[thick, blue] (\i, 1.1) -- ++(0, 0.9);
          \node[above] at (\i, 2) {$_2$};
        }
        \draw[blue, dashed, thick] (2, 0) -- ++(0, 2);
        \draw[blue, dashed, thick] (3, 0) -- ++(0, 2);
      \end{tikzpicture}
      = 0 \quad (\text{$2$-defects contracted} ) \\
    &\mathcal{O}_2 \cdot
      \begin{tikzpicture}[scale=0.5, baseline=(current bounding box.center)]
        \foreach \i in {0, 1, 2, 3}
        \draw[fill] (\i, 0) circle (2pt);
        \draw[thick, blue] (1, 0) to[out=-90, in=-90] (2, 0);
        \draw[thick, blue] (0, 0) -- ++(0, -1) node[below, black] {$_1$};
        \draw[thick, blue] (3, 0) -- ++(0, -1) node[below, black] {$_1$};
      \end{tikzpicture}
      =
      \begin{tikzpicture}[scale=0.5, baseline=(current bounding box.center)]
        \foreach \i in {0, 1, 2, 3}
        \draw[fill] (\i, 0) circle (2pt);
        \draw[thick, blue] (1, 0) to[out=-90, in=-90] (2, 0);
        \draw[thick, blue] (0, 0) -- ++(0, -1);
        \draw[thick, blue] (3, 0) -- ++(0, -1);
        \foreach \i in {0, 1} {
          \draw[fill] (\i, 0.9) circle (2pt);
          \draw[thick, blue] (\i, 0.9) -- ++(0, -0.9);
          \draw[fill] (\i, 1.1) circle (2pt);
          \draw[thick, blue] (\i, 1.1) -- ++(0, 0.9);
          \node[above] at (\i, 2) {$_2$};
        }
        \draw[blue, dashed, thick] (2, 0) -- ++(0, 2);
        \draw[blue, dashed, thick] (3, 0) -- ++(0, 2);
      \end{tikzpicture}
      =
      \begin{tikzpicture}[scale=0.5, baseline=(current bounding box.center)]
        \foreach \i in {0, 1, 2, 3}
        \draw[fill] (\i, 0) circle (2pt);
        \draw[thick, blue] (0, 0) -- ++(0, -1) node[below, black] {$_2$};
        \draw[thick, blue] (1, 0) -- ++(0, -1) node[below, black] {$_2$};
        \draw[thick, blue] (2, 0) -- ++(0, -1) node[below, black] {$_2$};
        \draw[thick, blue] (3, 0) -- ++(0, -1) node[below, black] {$_1$};
      \end{tikzpicture}
  \end{align}
\end{subequations}
At $t=M$, we impose the boundary condition $\langle \psi_3 |$, with labels $3$ on the defects. We project on $\langle \psi_3 |$ by using a generalisation of the bilinear form $\langle \cdot|\cdot \rangle$. Namely, we contract differently labelled defects with weight $1$, or create loops of weight $n$. If the outcome is anything else than a link pattern with one defect on each of the edges $x=1,2,\ldots,\ell_3$, we set the result to $0$.

If the transfer matrix contracts two defects carrying the same label, we set the result to $0$.
With these definitions, on the bottom half of the cylinder the link patterns all contain $\ell_1$ $1$-defects. In the top-half of the cylinder, link patterns contain between $|\ell_1-\ell_2|$ and $\ell_1 + \ell_2$ defects, with labels $1$ or $2$.

\subsubsection{$\langle V_{(r_1,0)} V_{(r_2,0)} V_{(r_3,0)} \rangle$ with enclosures}

In the presence of enclosures, we have to allow contractions of defects that carry the same label, as dictated by the combinatorial map. In order to allow the right contractions, we can no longer label defects simply by their vertices of origin: we now need to keep track of each defect. To do this, we use labels $1,2,\ldots,\ell_1$ for $1$-defects, and $\ell_1+1,\ldots,\ell_1+\ell_2$ for $2$-defects.

In the example $(\ell_1,\ell_2,\ell_3)=(4,1,1)$ of Figure (\ref{fig:cmaps}b), we may require the defect with labels $2$ and $4$ to connect with one another, forming an enclosure. On the other hand, the defects with labels $3$ and $5$ will also connect with one another, without forming an enclosure.

Then, in the combination~\eqref{eq:3}, only $Z_{123}$ involves enclosures, while the other factors do not. For consistency we must thus multiply $Z_{i0i}$ by combinatorial factors.

\subsubsection{One diagonal field: $\langle V_{(r_1,0)} V_{(0, s_2)} V_{(r_3,0)} \rangle$}

We now consider three-point functions of two leg fields, together with one diagonal field $\mathcal{O}_2$ at the point $(t,x)=(0,\frac12)$ in the middle of the cylinder. The problem is now to give a weight $w_2=2\cos(\pi s_2)$ to any closed loop that winds around $\mathcal{O}_2$.

\vspace{2mm}

\hspace*{-\parindent}
\begin{minipage}{0.72\textwidth}
  \setlength{\parindent}{1.5em}
  To do this, following \cite{ijs15}, we draw a seam line $\Sigma$ on the lattice between the fields $V_1$ and $V_2$, say the straight segment $(t,x)\in [-M,0]\times \{\frac12\}$. In a link pattern, we endow each arch with a binary variable $\sigma$ in $\mathbb{Z}_2 := \{ 0,1 \}$. The value of $\sigma$ denotes the parity of the number of times that the given arch has traversed $\Sigma$.

  When the action of $T$ joins two distinct arches with signs $\sigma$ and $\sigma'$, the sign of the resulting arch is $\sigma + \sigma'$. When the action of $T$ forms a loop by closing an arch with sign $\sigma$, we multiply with the weight $n$ if $\sigma = 0$, and $w_2$ if $\sigma = 1$.

  Due to the presence of defects originating from the top and bottom of the cylinder, closed loops must either be topologically trivial, or wind around $\mathcal{O}_2$: a $\mathbb{Z}_2$ variable is enough for distinguishing these possibilities.

  This construction also works for the factor $Z_{220}$ in Eq.~\eqref{eq:3}, with the seam $\Sigma$ detecting whether loops surround a non-trivial non-diagonal field at the bottom or middle of the cylinder. For $Z_{202}$, we should change the seam line to $(t,x)\in [-M,M]\times \{\frac12\}$.
\end{minipage}
\begin{minipage}{0.28\textwidth}
  \begin{align}
    \label{eq:18}
    \begin{tikzpicture}[scale=0.5, baseline=(current bounding box.center)]
      \draw[red] (0, 10) ellipse[x radius=2cm, y radius=0.2cm];
      \draw[red] (2, 0) arc[start angle=0, end angle=180, x radius=2cm, y radius=-0.2cm];
      \draw[red, dashed] (2, 0) arc[start angle=0, end angle=180, x radius=2cm, y radius=0.2cm]; 
      \fill[thick, red] (0, 4.8) circle (.1);
      \draw (-2, 0) -- ++ (0, 10);
      \draw (2, 0) -- ++ (0, 10);
      \draw[black!15!green, dashed] (0, -0.2) -- node[pos=0.25, right] {$\Sigma$}++ (0, 5);
      \draw[rounded corners=5pt, blue] (-1.3, 3.5) -- (-1.3, 3) -- (-1.3, 2.2) -- (0.5, 2.2) -- (0.5, 2.7) -- (-0.7, 3) -- (-0.7, 3.5) node[above=-3pt]{$_{\sigma = 0}$} ;
      \draw[rounded corners=7pt, blue] (0.8, 7) node[left]{$_{\sigma = 1}$} -- (-0.8, 5) -- (0, 4) -- (1.2, 5.5) -- (1.2, 7);
    \end{tikzpicture}
  \end{align}
\end{minipage}

\subsection{Three-point functions of fields with spin}
\label{subsec:3ptspin}

On the lattice, fields with nonzero spins are constructed by assigning phase factors to loop configurations. For a field $V_{(r,s)}$ with $\ell=2r$ legs, the phase factor is $e^{i\pi s\sigma}$, where  $\sigma \in \{0,1,\dots,\ell-1\}$ is an element of the cyclic group $\mathbb{Z}_\ell$ that permutes the $\ell$ legs, in the orientation of Figure \eqref{fig:cyl-lat}. In the case where the field $V_{(r, s)}$ is inserted at a boundary of the cylinder, the field $| V_{(r, s)} \rangle$ is in the module $\mathcal{W}_{(r, s)}$ of the $\uJTL_L(n)$ algebra defined in Section \ref{subsec:trm}.
In the case where the field $V_{(r, s)}$ is inserted in the middle 
The lattice 3-point function becomes a combination of $\ell_1\ell_2\ell_3$ terms,
\begin{align}
  \label{eq:7}
  Z_{123} &= \sum_{\sigma_i\in\mathbb{Z}_{\ell_i}} e^{i \pi \sum_i s_i \sigma_i} d_{\sigma_1, \sigma_2, \sigma_3}\ , \\
         &= \langle \psi_{(r_3, s_3)} | T^M \mathcal{O}_{(r_2, s_2)} T^M | \psi_{(r_1, s_1)} \rangle,
\end{align}
where the states $| \psi_{(r, s)} \rangle$ and the operator $\mathcal{O}_{(r, s)}$ are defined in \eqref{eq:psiOrs}, and $d_{\sigma_1, \sigma_2, \sigma_3}$ is a sum over configurations where the legs from the 3 vertices connect with one another in a specific manner. The difficulty in the transfer matrix approach is to compute separately the amplitudes $d_{\sigma_1, \sigma_2, \sigma_3}$ corresponding to all possible ways of connecting the legs together.

\subsubsection{Case of $\left< V_{(r_1, s_1)} V_{(r_2, s_2)} V_{(r_3, s_3)} \right>$, $r_i>0$, no enclosures}

Our transfer matrix algorithm computes $d_{\sigma_1, \sigma_2, \sigma_3}$ as
\begin{align}
  \label{eq:20}
  d_{\sigma_1, \sigma_2, \sigma_3} = \langle \psi_{\ell_3}^{\sigma_3} | T^M \mathcal{O}_{\ell_2}^{\sigma_2} T^M | \psi_{\ell_1}^{\sigma_1} \rangle \ .
\end{align}
We ensure that in $d_{0, 0, 0}$, the first $\frac12(\ell_1+\ell_2-\ell_3)$ legs coming from the bottom connect with legs coming from the middle insertion, while all other bottom legs connect to the last $\frac12(\ell_1+\ell_3-\ell_2)$ top legs, and by construction $d_{\sigma_1, \sigma_2, \sigma_3}$ is obtained from $d_{0, 0, 0}$ by cyclically permuting legs at each insertion. A different choice of reference configuration $d_{0, 0, 0}$ gives the same result for $Z_{123}$ up to an overall phase.
More precisely:
\begin{itemize}
\item
  At the bottom of the cylinder, we insert a state $| \psi_{\ell_1}^{\sigma_1} \rangle$, which we obtain from $| \psi_{\ell_1}^{\sigma_1=0} \rangle$ by cyclically shifting all defects labels to the right by $\sigma_1 \in \mathbb{Z}_{\ell_1}$. The state $| \psi_{\ell_1}^{\sigma_1=0} \rangle$ is the same as $| \psi_{\ell_1} \rangle$ \eqref{eq:14}, except that the legs have a marker $b$ to indicate that they come from the bottom insertion, and labels running from $1, \dots, \ell_1$ according to the orientation of \eqref{fig:cyl-lat}.
\item
  At the middle of the cylinder we insert an operator $\mathcal{O}_{\ell_2}^{\sigma_2}$ which inserts labelled legs. Labels for $\mathcal{O}_{\ell_2}^{\sigma_2}$ are obtained from those for $\mathcal{O}_{\ell_2}^{\sigma_2=0}$ by cylically permuting labels by $\sigma_2\in \mathbb{Z}_{\ell_2}$ clockwise, and $\mathcal{O}_{\ell_2}^{\sigma_2}$ is the same as $\mathcal{O}_{\ell_2}$ except legs now have a marker $m$ to indicate that they come from the middle insertion, and labels running from $1, \dots, \ell_2$.
  The action of $\mathcal{O}_2^{\sigma}$ and of the transfer matrix on link patterns on the top half cylinder is defined such that two defects can only be contracted together if they have different markers and identical labels.
\item
  At the top of the cylinder, we insert the state $\langle \psi_{\ell_3}^{\sigma_3} |$, where $| \psi_{\ell_3}^{\sigma_3} \rangle$ is obtained by cyclically shifting the $\ell_3$ legs of $| \psi_3^{\sigma_3=0} \rangle$ $\sigma_3$ times towards the left, where
  \begin{align}
    \label{fig:psi3spin}
    | (\psi_3^{\sigma_3 = 0})^{O(n)} \rangle &=
                                 \begin{tikzpicture}[scale=0.7, baseline=(current bounding box.center)]
                                   \foreach \i in {0, 1, 3, 4, 5, 7, 8, 9}
                                   \draw[fill] (\i, 0) circle (2pt);
                                   \node at (2, 0) {$\dots$};
                                   \foreach \i in {0, 1, 3, 4, 5, 7}
                                   \draw[thick, blue] (\i, 0) -- ++(0, -1);
                                   \node[below] at (0, -1) {$_{\substack{m \\ _{\ell_2-1}}}$};
                                   \node[below] at (1, -1) {$_{\substack{m \\ _{\ell_2-2}}}$};
                                   \node[below] at (3, -1) {$_{\substack{m\\\dots}}$};
                                   \node[below] at (4, -1) {$_{\substack{b \\ _{\ell_1-1}}}$};
                                   \node[below] at (5, -1) {$_{\substack{b \\ _{\ell_1-2}}}$};
                                   \node[below] at (7, -1) {$_{\substack{b\\\dots}}$};
                                   \node at (10, 0) {$\dots$};
                                   \draw [decorate,decoration={brace,amplitude=5pt}] (0, 5pt) -- node[midway, above, anchor=south]{$_{\substack{\frac12(\ell_2+\ell_3-\ell_1)\\\text{ defects}}}$}(3, 5pt);
                                   \draw [decorate,decoration={brace,amplitude=5pt}] (4, 5pt) -- node[midway, above, anchor=south]{$_{\substack{\frac12(\ell_1+\ell_3-\ell_2)\\\text{ defects}}}$}(7, 5pt);
                                 \end{tikzpicture}
                                 ,
  \end{align}
  and $| (\psi_3^{\sigma_3=0})^{PSU(n)} \rangle$ is the same, except we replace empty sites by arches like in \eqref{eq:14}. In figure \eqref{fig:psi3spin}, each leg carries a marker $b$ or $m$, and below a label.
\end{itemize}
With this construction, the lattice three-point function takes the form of \eqref{eq:7} provided
\begin{align}
  \label{eq:psiOrs}
  | \psi_{(r, s)} \rangle = \sum_{\sigma} e^{i \pi s \sigma} | \psi_r^{\sigma} \rangle, \\
  \mathcal{O}_{(r, s)} = \sum_{\sigma} e^{i \pi s \sigma} \mathcal{O}_{\ell=2r}^{\sigma}.
\end{align}

\subsubsection{Cases involving enclosures or diagonal fields}
These cases can be treated using a mixture of the transfer matrix algorithms defined in \ref{subsec:3ptspinless} and \ref{subsec:3ptspin}. Having not implemented them numerically, we refrain from describing them explicitly.

\subsection{Numerical implementation}

Numerically, we define the transfer matrix $T$ as a product of sparse matrices that each propagate through a single vertex. We represent link patterns by unique integer keys which we store in a hash table, a data structure that allows for $O(1)$ lookup and insertion. Implementations of the above constructions in \texttt{C} and \texttt{C++} are available in a public repository \cite{roux25}. The main subtleties are:
\begin{itemize}
\item Before applying the boundary state $|\psi_3\rangle$ at the top of the cylinder, the states $|\sigma_t\rangle$ tend to be dominated by the largest eigenvalue $\Lambda$ of $T$. However, $\Lambda$ decreases with the number $\ell$ of defects. On the upper half of the cylinder we have link patterns with $|\ell_1-\ell_2|\leq \ell\leq \ell_1+\ell_2$, while applying $|\psi_3\rangle$ kills all link patterns with $\ell\neq \ell_3$. If $|\ell_1-\ell_2|<\ell_3$, there are link patterns with $\ell<\ell_3$, whose contributions dominate those of link patterns with $\ell =\ell_3$. As a result, the latter can become zero in computations with finite numerical precision. To avoid this, we eliminate all patterns with $\ell<\ell_3$, as soon as they are produced when applying $T$ in the upper half of the cylinder.
\item Moreover, in the sum over diagrams \eqref{eq:7} there are cancellations between the dominant terms corresponding to the eigenvalue of the spinless field. After these cancellations, the remaining dominant contributions are those  of the fields with spin. We are thus forced to use arbitrary-precision arithmetics in order for the subdominant contributions not to be killed by numerical rounding. For efficiency, we independently compute the $\ell_1\ell_2$ vectors $T^M \mathcal{O}_2^{\sigma_2} T^M | \psi_1^{\sigma_1} \rangle$, and we then just need to project each of them on the $\ell_3$ vectors $| \psi_3^{\sigma_3} \rangle$ to get all of the $\ell_1\ell_2\ell_3$ diagrams $d_{\sigma_1, \sigma_2, \sigma_3}$.
\end{itemize}

\subsection{Critical limit}\label{subsec:limit}

In the limit $M\to \infty$ of a very long cylinder, $T^M$ is dominated by the largest eigenvalue of the transfer matrix $T$. Let us call $\Lambda_{(r, s)}$ the largest eigenvalue of the action $T_{(r, s)}$ of $T$ in the $\uJTL_L(n)$ module $\mathcal{W}_i$ corresponding to the field $V_{(r, s)}$, and call $| V_{(r, s)}^j \rangle, j = 1, \dots, \dim \mathcal{W}_{(r, s)}$ the corresponding eigenvectors, ordered by decreasing (in norm) eigenvalue. When the leading eigenvalue of $T_{(r, s)}$ is not degenerate, we denote $| V_{(r, s)} \rangle = | V_{(r, s)}^1 \rangle$ the corresponding eigenvectors.

The eigenvectors are never computed numerically, because they are very complicated: in practice we work with simpler states $\psi_i$, which belong to the same modules, see (\ref{eq:13},~\ref{eq:14}). Introducing the eigenvectors nevertheless allows us to prove that three-point functions have finite limits.

To obtain the simplest expressions, we orthonormalise the eigenvectors with respect to the bilinear form $\langle \cdot | \cdot \rangle$ \eqref{eq:psi0}, so that $\langle V_{(r_1, s_1)}^{j_1} | V_{(r_2, s_2)}^{j_2} \rangle = \delta_{r1,r_2}\delta_{s_1,s_2}\delta_{j_1,j_2}$.

At generic central charges, the leading eigenvalues of $T_{(r, s)}$ are non-degenerate in any module $\mathcal{W}_{(r, s\neq 1)}$. Starting with the expression~\eqref{eq:17} and decomposing $\psi_i$ over the basis of eigenvectors we thus find in the spinless case:
\begin{align}
  \label{eq:Z123limM}
  Z_{123} \underset{M\to\infty}{\sim} \Lambda_1^M\Lambda_3^M \langle \psi_3 | V_3 \rangle \langle V_3 | \mathcal{O}_2 | V_1 \rangle \langle V_1 | \psi_1 \rangle\ .
\end{align}
In the case of fields with spins, $Z_{123}$ is a linear combination of coefficients of the type $\langle \psi_3 | T^M \mathcal{O}_2 T^M | \psi_1 \rangle$, \eqref{eq:7}. Because the bottom and top legs in $Z_{123}$ carry representations $e^{i\pi s_1}, e^{i \pi s_3}$ of $\mathbb{Z}_{\ell_1}$, $\mathbb{Z}_{\ell_3}$ respectively, the propagation in the bottom and top halves of the cylinder effectively projects on the module $\mathcal{W}_{(r, s)}$ described in section \ref{subsec:case_s=1}, so that
\begin{align}
  \label{eq:6}
  Z_{123} \underset{M\to\infty}{\sim} \Lambda_1^M\Lambda_3^M \langle \psi_3 | V_3 \rangle \left( \sum_{\sigma_2} e^{i \pi \sigma_2 s_2} \langle V_3 | \mathcal{O}^{\sigma_2}_2 | V_1 \rangle  \right)\langle V_1 | \psi_1 \rangle \ .
\end{align}
Now, if we compute the ratio $C_{123}(M,L)$~\eqref{eq:3}, the factors $\Lambda_i^M$ all cancel out, which proves that
\begin{align}
  \lim_{M \to \infty} C_{123} < \infty.
\end{align}
Using the symmetry of the bilinear form $\langle \cdot | \cdot \rangle$ the expression for $C_{123}$ simplifies to
\begin{equation}
  \label{eq:C123spin}
  C_{1 2 3}(L) = \frac{\sum_{\sigma_2} e^{i \pi \sigma_2 s_2} \langle V_3 | \mathcal{O}^{\sigma_2}_2 | V_1 \rangle}{\sum_{\sigma_2}e^{i \pi \sigma_2 s_2}\langle V_0 | \mathcal{O}^{\sigma_2}_2 | V_2 \rangle} \times \left( \prod_{i \in \{1, 3\}}  \frac{\langle V_i | \psi_i \rangle}{\sqrt{\langle V_i | \psi_i \rangle^2}} \right) \left( \prod_{i \in \{0, 2\}}  \frac{\sqrt{\langle V_i | \psi_i \rangle^2}}{\langle V_i | \psi_i \rangle} \right)\ ,
\end{equation}
which proves that $C_{123}(L)$ is independent of the choice of states $| \psi_i \rangle$ up to a sign, and of the normalisation of $\mathcal{O}_2$.
The sign ambiguity disappears in the spinless case provided the loop weight $n$ is positive. Indeed, in this case the transfer matrix has non-negative coefficients, so the Perron--Frobenius theorem proves that its leading eigenvalues are non-degenerate, and that the corresponding eigenvectors $V_i$ can be chosen to have real, non-negative elements in the basis of link patterns. In particular, $\langle V_i|\psi_i \rangle > 0$, so that
\begin{align}
  C_{1 2 3}(L) = \frac{\langle V_3 | \mathcal{O}_2 | V_1 \rangle}{\langle V_0 | \mathcal{O}_2 | V_2 \rangle}.
  \label{eq:C123}
\end{align}
Numerically, we find that $C_{123}(M,L)$ is a number of order unity, which stabilises when $M$ becomes large. Working with standard 16-digit floating-point arithmetics, we find that taking $M=20 L$ is sufficient for reaching the large-$M$ limit.
Taking the limit $L\to\infty$ is much more difficult, and cannot be tackled by standard algebraic methods. We detail in section~\ref{subsec:101010} how we perform it numerically.

The expression \eqref{eq:C123spin} provides a direct generalisation of \eqref{eq:3} to the case of primary fields $V_{(r, s)} = V_{(r, s_0+2n)}$ corresponding to subleading eigenstates of the transfer matrix. We conjecture that \eqref{eq:2} still holds in this case, but we do not give numerical evidence for this, given the difficulty of numerically computing subleading eigenstates for large system sizes.

\subsubsection{Worked out example: $C^{PSU(n)}_{(1, 0)(1, 0)(1, 0)}(L=4)$}

We now illustrate the construction by detailing the computation of $C^{PSU(n)}_{(1, 0)(1, 0)(1, 0)}(L=4)$ in size $L=4$. We restrict for simplicity to the $PSU(n)$ model because the corresponding modules have small dimensions: the modules with 0 and 2 defects have bases
\begin{align}
  \label{eq:1}
  \mathcal{W}_0: \qquad
  &\begin{tikzpicture}[scale=0.5, baseline=(base)]
    \coordinate (base) at (0, -.5);
    \foreach \i in {1, 2, 3, 4}
    \draw[fill] (\i, 0) circle (2pt);
    \draw[thick, blue] (1, 0) to[out=-90, in=-90] (2, 0);
    \draw[thick, blue] (3, 0) to[out=-90, in=-90] (4, 0);
  \end{tikzpicture} \hspace{0.3cm}
    \hspace{1cm}
    \begin{tikzpicture}[scale=0.5, baseline=(base)]
      \coordinate (base) at (0, -.5);
      \foreach \i in {1, 2, 3, 4}
      \draw[fill] (\i, 0) circle (2pt);
      \draw[thick, blue] (1, 0) to[out=-90, in=-90] (4, 0);
      \draw[thick, blue] (2, 0) to[out=-50, in=-130] (3, 0);
    \end{tikzpicture} \hspace{0.3cm} \\
  \mathcal{W}_2: \qquad
  &\begin{tikzpicture}[scale=0.5, baseline=(current bounding box.center)]
    \foreach \i in {0, 1, 2, 3}
    \draw[fill] (\i, 0) circle (2pt);
    \draw[thick, blue] (2, 0) to[out=-90, in=-90] (3, 0);
    \draw[thick, blue] (0, 0) -- ++(0, -1); 
    \draw[thick, blue] (1, 0) -- ++(0, -1); 
  \end{tikzpicture} \hspace{1cm}
    \begin{tikzpicture}[scale=0.5, baseline=(current bounding box.center)]
      \foreach \i in {0, 1, 2, 3}
      \draw[fill] (\i, 0) circle (2pt);
      \draw[thick, blue] (0, 0) to[out=-90, in=0] (-0.5, -0.5);
      \draw[thick, blue] (3, 0) to[out=-90, in=180] (3.5, -0.5);
      \draw[thick, blue] (1, 0) -- ++(0, -1); 
      \draw[thick, blue] (2, 0) -- ++(0, -1); 
    \end{tikzpicture} \hspace{1cm}
    \begin{tikzpicture}[scale=0.5, baseline=(current bounding box.center)]
      \foreach \i in {0, 1, 2, 3}
      \draw[fill] (\i, 0) circle (2pt);
      \draw[thick, blue] (0, 0) to[out=-90, in=-90] (1, 0);
      \draw[thick, blue] (2, 0) -- ++(0, -1); 
      \draw[thick, blue] (3, 0) -- ++(0, -1); 
    \end{tikzpicture} \hspace{1cm}
    \begin{tikzpicture}[scale=0.5, baseline=(current bounding box.center)]
      \foreach \i in {0, 1, 2, 3}
      \draw[fill] (\i, 0) circle (2pt);
      \draw[thick, blue] (1, 0) to[out=-90, in=-90] (2, 0);
      \draw[thick, blue] (0, 0) -- ++(0, -1); 
      \draw[thick, blue] (3, 0) -- ++(0, -1); 
    \end{tikzpicture} \hspace{1cm}
\end{align}
whence $\dim \mathcal{W}_0 = 2$ and $\dim \mathcal{W}_2 = 4$. (In the $O(n)$ model, $\dim \mathcal{W}_0 = 9$ and $\dim \mathcal{W}_2 = 10$.)
In these bases, we compute the matrix elements $\langle \psi |\mathcal{O}_2 | \psi' \rangle$ by the diagrammatic rules described in section~\ref{subsec:3ptspinless}. Note that, since $\mathcal{O}_2$ inserts defects, it can have non-zero elements between states in modules with different numbers of legs. For instance,
\begin{align}
  \label{eq:4}
  &\left\langle\hspace{0.3cm}
    \begin{tikzpicture}[scale=0.5, baseline=(base)]
      \coordinate (base) at (0, -.5);
      \foreach \i in {1, 2, 3, 4}
      \draw[fill] (\i, 0) circle (2pt);
      \draw[thick, blue] (1, 0) - ++(0, -1);
      \draw[thick, blue] (2, 0) - ++(0, -1);
      \draw[thick, blue] (3, 0) to[out=-90, in=-90] (4, 0);
    \end{tikzpicture} \hspace{0.3cm}
    \middle| \mathcal{O}_2 \middle| \hspace{0.3cm}
    \begin{tikzpicture}[scale=0.5, baseline=(base)]
      \coordinate (base) at (0, -.5);
      \foreach \i in {1, 2, 3, 4}
      \draw[fill] (\i, 0) circle (2pt);
      \draw[thick, blue] (1, 0) to[out=-90, in=-90] (4, 0);
      \draw[thick, blue] (2, 0) to[out=-50, in=-130] (3, 0);
    \end{tikzpicture} \hspace{0.3cm}
    \right\rangle =
    \begin{tikzpicture}[scale=0.5, baseline=(current bounding box.center)]
      \foreach \i in {2, 3, 4}
      \draw[fill] (\i, 0) circle (2pt);
      \draw[fill] (1, 0.1) circle (2pt);
      \draw[fill] (1, -0.1) circle (2pt);
      \draw[thick, blue] (1, 0.1) -- ++(0, 0.9);
      \draw[thick, blue] (2, 0) - ++(0, 1);
      \draw[thick, blue] (3, 0) to[out=90, in=90] (4, 0);
      \draw[thick, blue] (1, -0.1) to[out=-90, in=-90] (4, 0);
      \draw[thick, blue] (2, 0) to[out=-50, in=-130] (3, 0);
    \end{tikzpicture} \hspace{0.3cm} = 1 \\
  &\left\langle\hspace{0.3cm}
    \begin{tikzpicture}[scale=0.5, baseline=(base)]
      \coordinate (base) at (0, -.5);
      \foreach \i in {1, 2, 3, 4}
      \draw[fill] (\i, 0) circle (2pt);
      \draw[thick, blue] (1, 0) - ++(0, -1);
      \draw[thick, blue] (2, 0) - ++(0, -1);
      \draw[thick, blue] (3, 0) to[out=-90, in=-90] (4, 0);
    \end{tikzpicture} \hspace{0.3cm}
    \middle| \mathcal{O}_2 \middle| \hspace{0.3cm}
    \begin{tikzpicture}[scale=0.5, baseline=(base)]
      \coordinate (base) at (0, -.5);
      \foreach \i in {1, 2, 3, 4}
      \draw[fill] (\i, 0) circle (2pt);
      \draw[thick, blue] (1, 0) - ++(0, -1);
      \draw[thick, blue] (2, 0) - ++(0, -1);
      \draw[thick, blue] (3, 0) to[out=-90, in=-90] (4, 0);
    \end{tikzpicture} \hspace{0.3cm}
    \right\rangle =
    \begin{tikzpicture}[scale=0.5, baseline=(current bounding box.center)]
      \foreach \i in {2, 3, 4}
      \draw[fill] (\i, 0) circle (2pt);
      \draw[fill] (1, 0.1) circle (2pt);
      \draw[fill] (1, -0.1) circle (2pt);
      \draw[thick, blue] (1, 0.2) -- ++(0, 0.8);
      \draw[thick, blue] (2, 0) - ++(0, 1);
      \draw[thick, blue] (3, 0) to[out=90, in=90] (4, 0);
      \draw[thick, blue] (1, -0.2) -- ++(0, -0.8);
      \draw[thick, blue] (2, 0) - ++(0, -1);
      \draw[thick, blue] (3, 0) to[out=-90, in=-90] (4, 0);
    \end{tikzpicture} \hspace{0.3cm} = n \\
  &\left\langle\hspace{0.3cm}
    \begin{tikzpicture}[scale=0.5, baseline=(base)]
      \coordinate (base) at (0, -.5);
      \foreach \i in {1, 2, 3, 4}
      \draw[fill] (\i, 0) circle (2pt);
      \draw[thick, blue] (3, 0) - ++(0, -1);
      \draw[thick, blue] (4, 0) - ++(0, -1);
      \draw[thick, blue] (1, 0) to[out=-90, in=-90] (2, 0);
    \end{tikzpicture} \hspace{0.3cm}
    \middle| \mathcal{O}_2 \middle| \hspace{0.3cm}
    \begin{tikzpicture}[scale=0.5, baseline=(base)]
      \coordinate (base) at (0, -.5);
      \foreach \i in {1, 2, 3, 4}
      \draw[fill] (\i, 0) circle (2pt);
      \draw[thick, blue] (3, 0) - ++(0, -1);
      \draw[thick, blue] (4, 0) - ++(0, -1);
      \draw[thick, blue] (1, 0) to[out=-90, in=-90] (2, 0);
    \end{tikzpicture} \hspace{0.3cm}
    \right\rangle =
    \begin{tikzpicture}[scale=0.5, baseline=(current bounding box.center)]
      \foreach \i in {2, 3, 4}
      \draw[fill] (\i, 0) circle (2pt);
      \draw[fill] (1, 0.2) circle (2pt);
      \draw[fill] (1, -0.2) circle (2pt);
      \draw[thick, blue] (3, 0) - ++(0, -1);
      \draw[thick, blue] (4, 0) - ++(0, -1);
      \draw[thick, blue] (1, -0.2) to[out=-90, in=-90] (2, 0);
      \draw[thick, blue] (3, 0) - ++(0, 1);
      \draw[thick, blue] (4, 0) - ++(0, 1);
      \draw[thick, blue] (1, 0.2) to[out=90, in=90] (2, 0);
    \end{tikzpicture} \hspace{0.3cm} = 0 \quad (\text{2-defects contracted}).
\end{align}
Using this, we compute the matrix elements of $\mathcal{O}_2$ between 0 and 2 leg states:
\begin{align}
  \label{eq:8}
  \langle \mathcal{W}_2 | \mathcal{O}_2 | \mathcal{W}_2 \rangle = \begin{pmatrix}
    n & 1 & 0 & 1 \\
    1 & 0 & 1 & 0 \\
    0 & 1 & 0 & 1 \\
    1 & 0 & 1 & n
  \end{pmatrix} \qquad \langle \mathcal{W}_2 | \mathcal{O}_2 | \mathcal{W}_0 \rangle = \begin{pmatrix}
    n & 1 \\ 1 & 0 \\ 0 & 1 \\ 1 & n
  \end{pmatrix}.
\end{align}
By explicitly diagonalising the matrices
\begin{align}
  \label{eq:11}
  T_2 = \begin{pmatrix}
    2 & 2+n & 2+n & 2+n \\
    2+n & 2 & 2+n & 2+n \\
    2+n & 2+n & 2 & 2+n \\
    2+n & 2+n & 2+n & 2 \\
  \end{pmatrix}, \hspace{1cm} T_0 = \begin{pmatrix}
    4+3n     & 4+4n+n^2 \\
    4+4n+n^2 & 4+3n
  \end{pmatrix}
\end{align}
of the action of $T$ on $\mathcal{W}_2$ and $\mathcal{W}_0$, we arrive at
\begin{align}
  \label{eq:10}
  C_{(1, 0)(1, 0)(1, 0)}^{PSU(n)}(L=4) = \frac{(4+n)}{(2+n)}\sqrt{\frac{n(1+n)}{2(2+n)}},
\end{align}
which exactly fits our numerical results in size 4.

\section{Results for spinless fields}\label{sec:res}

We now present our numerical results for the three-point structure constants $C_{123}(L)$, discuss how they can be extrapolated to the limit $L\to\infty$, and compare with the analytic conjecture $\omega_{123}$.

We have tested all cases with no enclosures or diagonal fields with $r_i \leq 2$ and $\sum r_i \leq 5$. For the cases with diagonal fields or enclosures, we present our results in sections \ref{subsec:caser0r0r0} and \ref{subsec:case100s10}.
All of the numerical results can be found in the repository \cite{roux25}.

\subsection{Case of $\langle V_{(1,0)} V_{(1,0)} V_{(1,0)} \rangle$}\label{subsec:101010}

We first give a detailed presentation of the methodology for the case $\langle V_{(1,0)} V_{(1,0)} V_{(1,0)} \rangle$ of three spinless two-leg fields. The correlation function exists both in the $O(n)$ and the $PSU(n)$ loop models.

\begin{figure}[htbp]
  \centering
  \includegraphics[width=0.8\textwidth]{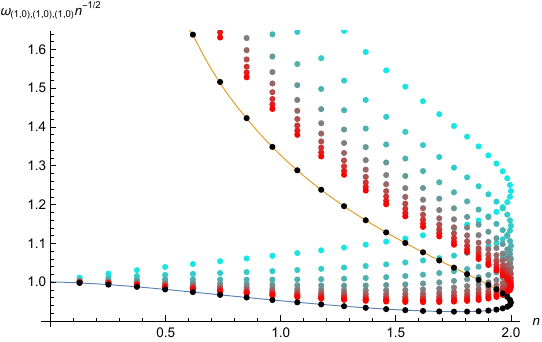}
  \caption{Finite-size results for $\frac{C_{(1,0)(1,0)(1,0)}(L)}{\sqrt n}$ in the $O(n)$ model, as functions of $n$.
    The continuous line plots $\frac{\omega_{(1,0)(1,0)(1,0)}}{\sqrt n}$, with
    a lower branch (blue) for the dense phase, and an upper branch (orange) for the dilute phase. The coloured points show numerical results for lattices of finite sizes $L=4$ (cyan) to $L=13$  (red). The black points are $L \to \infty$ extrapolations from $8 \le L \le 13$.}
  \label{fig:3ptOn222}
\end{figure}

Figure~\ref{fig:3ptOn222} shows the results for $C_{(1,0)(1,0)(1,0)}(L)$ obtained in the $O(n)$ model for sizes in the interval $4 \le L \le 13$.
The values of $n$ shown correspond to varying the parameter $\beta^2$ from $0.52$
to $1.48$ in steps of $0.02$.
This gives access to both the dense phase (lower part of the figure) and the dilute phase (upper part).

One first feature is that finite-size effects are strong, except when $n \to 0$ in the dense phase, and that they become more and more pronounced as we move into the
dilute phase. These effects will be even stronger when we consider higher values of the number of legs $\ell_i$.
We can nevertheless obtain precise final results by fitting, for each $\beta^2$, all data points with $L$ in some interval,
$L_{\mathrm{min}} \le L \le L_{\mathrm{max}}$, to a polynomial in the variable $1/L$ of degree $L_{\mathrm{max}} - L_{\mathrm{min}}$.
The black points in Figure~\ref{fig:3ptOn222} show the constant term in this polynomial fit for the choice $8 \le L \le 13$. Using all
data points ($4 \le L \le 13$) would give an even better result, which is however graphically indistinguishable from the one shown. Continous curves show the analytic predictions.
\begin{figure}[htbp]
  \centering
  \includegraphics[width=0.8\textwidth]{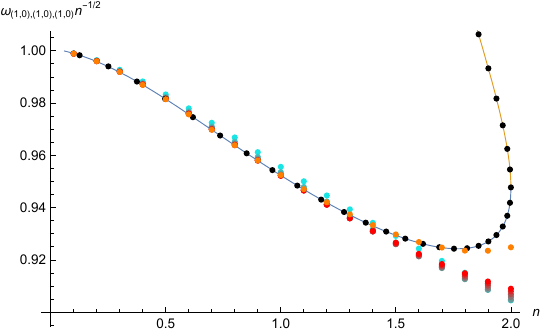}
  \caption{Finite-size results for $\frac{C_{(1,0)(1,0)(1,0)}(L)}{\sqrt n}$ in the $PSU(n)$ model, as functions of $n$.
    The coloured points show the finite even sizes from $L=6$ (cyan) to $L=18$ (red).
    Orange points give the $L \to \infty$ extrapolation, black points the corresponding $O(n)$ result for comparison.}
  \label{fig:3ptPotts222}
\end{figure}

Figure~\ref{fig:3ptPotts222} gives the corresponding results for the $PSU(n)$ loop model, for even sizes in the interval $6 \le L \le 18$. The data points shown correspond to varying $n$ from $0.1$ to $2$ in steps of $0.1$.
The finite-size effects are much smaller than in the $O(n)$ model, but actually the extrapolations (shown as orange points) turn out to be less precise, in particular close to $n=2$. Comparison with the $O(n)$ model (black points) nonetheless provides compelling evidence that continuum-limit results for both models are identical. This conclusion is corroborated for higher (even) numbers of legs.

Therefore, the $O(n)$ model has a number of advantages over the $PSU(n)$ model, as far as the study of three-point structure constants is concerned:
\begin{enumerate}
\item it gives access to both the dense and the dilute phase;
\item it imposes no parity constraints on $L$, or on the number of legs $\ell$;
\item it experiences no particular problems as $n \to 2$, whereas in the $PSU(n)$ model a dangerously irrelevant field makes numerical simulations more difficult \cite{ss97}.
\item it provides more precise extrapolations for the same computational effort, even though the sizes $L$ that can be reached in the transfer-matrix computations are smaller.
\end{enumerate}
For those reasons we present all subsequent numerical results for the $O(n)$ model only.

\begin{figure}[htbp]
  \centering
  \includegraphics[width=0.5\textwidth]{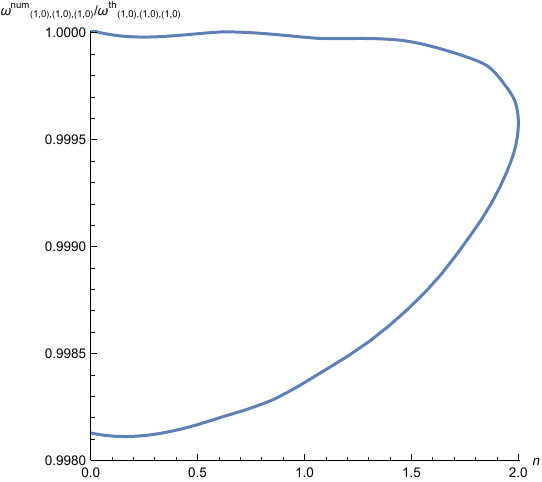}
  \caption{Ratio between the extrapolation $C_{(1,0)(1,0)(1,0)}(\infty)$ of the $O(n)$ loop-model results and $\omega_{(1,0),(1,0),(1,0)}$.}
  \label{fig:3ptOnRatio222}
\end{figure}

We finally give a high-precision comparison between the numerical and analytic results.
Figure~\ref{fig:3ptOnRatio222} presents the ratio between the numerical values for $C_{(1,0)(1,0)(1,0)}(\infty)$, here extrapolated from all sizes
$4 \le L \le 13$, and the analytic formula for $\omega_{(1,0),(1,0),(1,0)}$. The curve is obtained by interpolation from the data
points that were computed numerically, and the tiny ripples can be attributed to the interpolation scheme. The agreement is
excellent throughout the dense phase ($\tfrac12 \le \beta^2\le 1$) and still quite satisfactory in the dilute phase
($1 \le \beta^2 \le \frac{3}{2}$). For instance, when $n=1$, the ratio is $0.99998$ in the dense phase (percolation)
and $0.998$ in the dilute phase (Ising model).

\subsection{Case of $\langle V_{(\frac12,0)} V_{(1,0)} V_{(\frac32,0)} \rangle$}\label{subsec:V12010320}

In this case, the numbers of legs $1,2,3$ of the three fields are all different, allowing us to test permutation symmetry. While the 3-point function must be invariant under permutations of the three fields, the finite-size quantities $C_{123}(L)$, $C_{132}(L)$ and $C_{213}(L)$ are different, because the field at the middle of the cylinder is treated differently from the boundary states at the top and bottom.
\begin{figure}[htbp]
  \centering
  \includegraphics[width=0.48\textwidth]{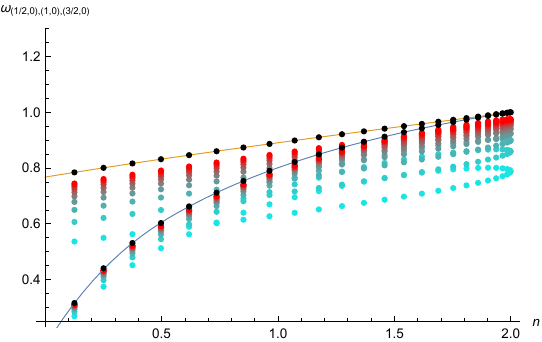} \quad
  \includegraphics[width=0.48\textwidth]{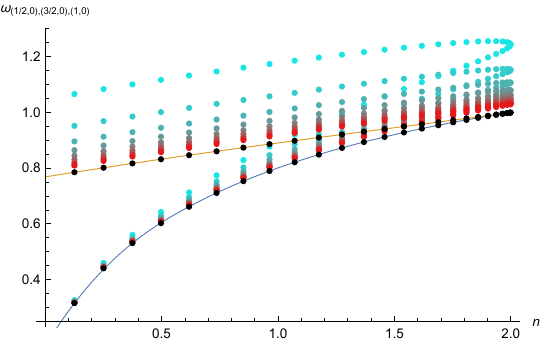}

  \medskip

  \includegraphics[width=0.48\textwidth]{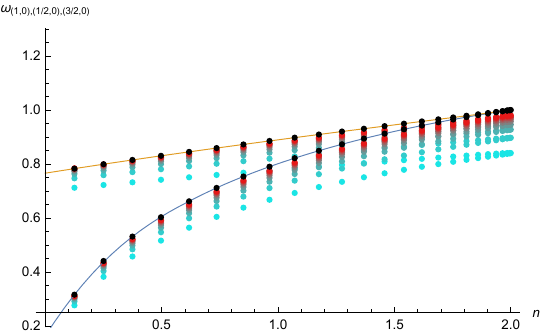}
  \caption{Comparison between $C_{123}(L)$ (upper left panel), $C_{132}(L)$ (upper right panel), and $C_{213}(L)$ (lower left panel). The black points show the $L\to \infty$ extrapolations.}
  \label{fig:3ptOn123}
\end{figure}

Figure~\ref{fig:3ptOn123} shows those three quantities, with the exact same conventions as those used in Figure~\ref{fig:3ptOn222}.
The finite-size effects have quite different trends. Indeed, $C_{123}(L)$ and $C_{213}(L)$ are both monotonically increasing functions of $L$, for any value of $n$ on both the dense and dilute branches, but the variations are stronger for $C_{123}(L)$. On the other hand, $C_{132}(L)$ is monotonically decreasing and its variations are even stronger. Remarkably, the $L \to \infty$ extrapolations of all three quantities, shown as black points in Figure~\ref{fig:3ptOn123}, are fully consistent with the same continuum limit and in excellent agreement with the analytic formula.

\subsection{Case of $\langle V_{(r_1,0)} V_{(r_2,0)} V_{(r_3,0)} \rangle$}\label{subsec:caser0r0r0}

\begin{figure}[htbp]
  \centering
  \includegraphics[width=0.31\textwidth]{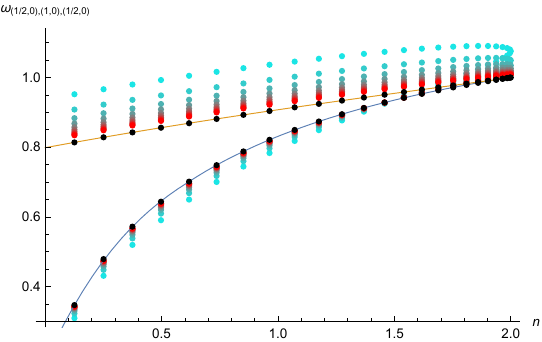} \quad
  \includegraphics[width=0.31\textwidth]{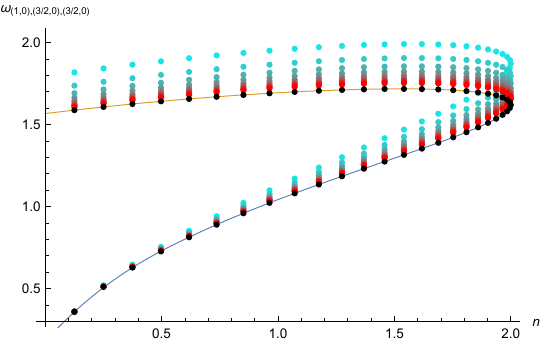} \quad
  \includegraphics[width=0.31\textwidth]{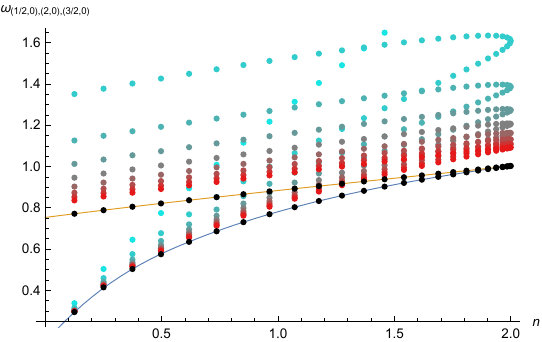} \quad
  
  \medskip
  
  \includegraphics[width=0.31\textwidth]{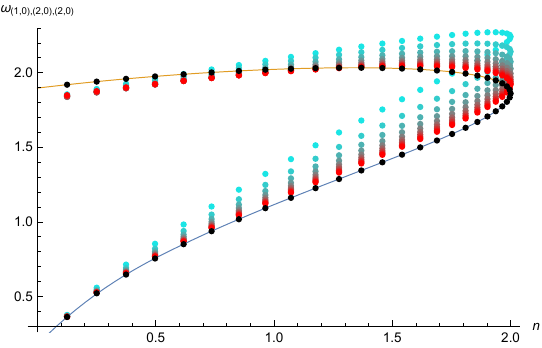}
  \includegraphics[width=0.31\textwidth]{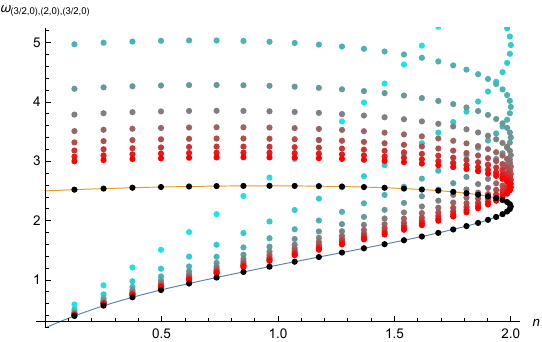} \quad
  \includegraphics[width=0.31\textwidth]{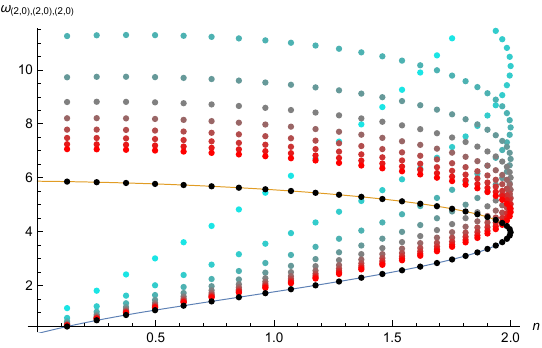} \quad
  
  \caption{Plots of $C_{123}(L)$ for the $O(n)$ loop-model results, their $L\to \infty$ extrapolations and the analytic formula $\omega_{(r_1,0),(r_2,0),(r_3,0)}$, for the cases
    $(r_1, r_2, r_3) = (\frac12,1,\frac12), (1,\frac32,\frac32), (\frac12,2,\frac32)$ (top row, from left to right), and $(1, 2, 2), (\frac32,2,\frac32), (2,2,2)$ (bottom row).}
  \label{fig:3ptOn6}
\end{figure}

In cases without enclosures, we have computed 6 different three-point functions, on top of the two examples of~\ref{subsec:101010} and \ref{subsec:V12010320}.
In Figure~\ref{fig:3ptOn6} we show these quantities and compare their $L \to \infty$ extrapolations with the analytic result $\omega_{(r_1,0),(r_2,0),(r_3,0)}$. The conventions for those plots are the
same as in Figure~\ref{fig:3ptOn222}. The agreement with the analytic result is excellent for small values of the numbers of legs $\ell_i=2r_i$, with the ratio being equal to unity to within $10^{-3}$, uniformly in $n$ for both the dense and dilute phases. For higher numbers of legs, the agreement is of order $10^{-2}$ uniformly, which is still satisfactory.

\begin{figure}[htbp]
  \centering
  \includegraphics[width=0.47\textwidth]{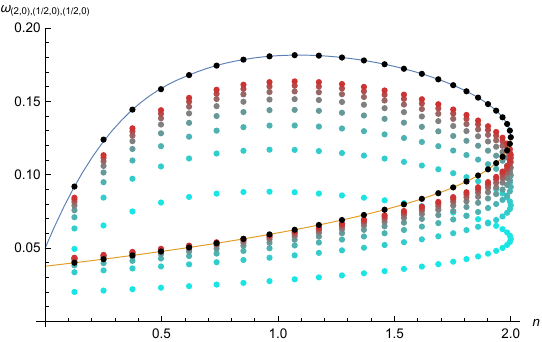} \quad
  \includegraphics[width=0.47\textwidth]{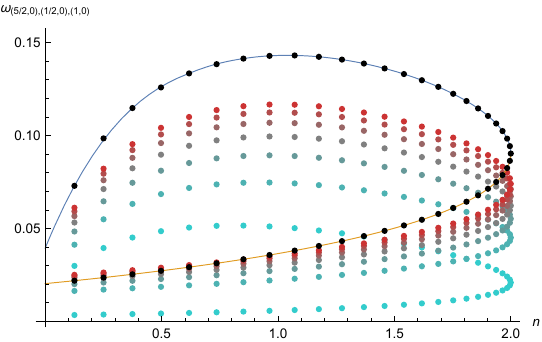}
  \caption{Plots of $C_{123}(L)$ for the $O(n)$ loop-model results, their $L\to \infty$ extrapolations and the analytic formula $\omega_{(r_1,0),(r_2,0),(r_3,0)}$, for two cases with enclosures:
    $(r_1, r_2, r_3) = (2,\frac12,\frac12)$ and $(\frac52,\frac12,1)$.}
  \label{fig:3ptOn411}
\end{figure}

For the cases with enclosures we made computations for the cases $(2r_1,2r_2,2r_3) = (4,1,1)$ and $(5,1,2)$, obtaining again
fine agreement with the analytic results, with a relative precision of the order $10^{-2}$ or better for all values of $n$.
This is shown in the usual way in Figure~\ref{fig:3ptOn411}.

\subsection{Case of $\langle V_{(1,0)} V_{(0, s)} V_{(1,0)} \rangle$}\label{subsec:case100s10}

\begin{figure}[htbp]
  \centering
  \includegraphics[width=0.47\textwidth]{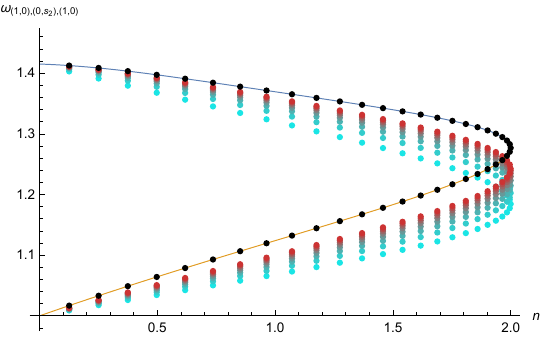} \quad
  \includegraphics[width=0.47\textwidth]{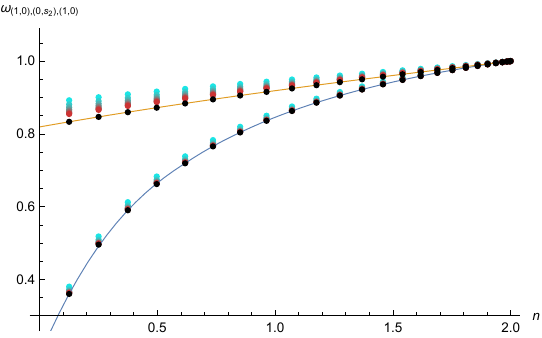}
  \caption{Extrapolated transfer-matrix results for the $O(n)$ loop-model structure constants for the correlation function $\langle V_{(1,0)} V_{(0, s)} V_{(1,0)} \rangle$ with $s=\frac12$ (left panel) and $s=0$ (right panel).}
  \label{fig:3ptOnRatio22}
\end{figure}

In the case of one diagonal and two non-diagonal fields, Figure~\ref{fig:3ptOnRatio22} displays the numerical data with their extrapolations as well as the analytic results for the case $\langle V_{(1,0)} V_{(0, s)} V_{(1,0)} \rangle$. The modified weight $w=2\cos(\pi s)$ of loops encircling the $V_{(0, s)}$ field are chosen as $w=0$ and $w=2$, respectively. In both cases the agreement is to within a relative error of the order $10^{-3}$. Other three-point functions with one diagonal field were examined as well, and again we found a convincing agreement with the analytic results.

\section{Case of fields with spin}\label{sec:nonzerospin}

In the case of fields with spin, we have tested 43 cases without enclosures nor diagonal fields, and with
\begin{align}
  (r_1, r_2, r_3) \in \{(1, 1, 1), (\tfrac12, 1, \tfrac32), (\tfrac32, 1, \tfrac32), (2, \tfrac12, \tfrac32), (2, 1, 1),  (2, 1, 2)\},
\end{align}
and all possible values of $s_i \in (-1, 1]$ with $(s_1, s_2, s_3) \neq (0, 0, 0)$, taking into account the parity symmetry $(s_1, s_2, s_3) \to (-s_1, -s_2, -s_3)$.

\subsection{Cases of $\left< V_{(r_1, s_1)} V_{(r_2, s_2)} V_{(r_3, s_3)} \right>$ with $s_i \neq 1$}

\begin{figure}[htbp]
  \centering
  \includegraphics[width=0.45\textwidth]{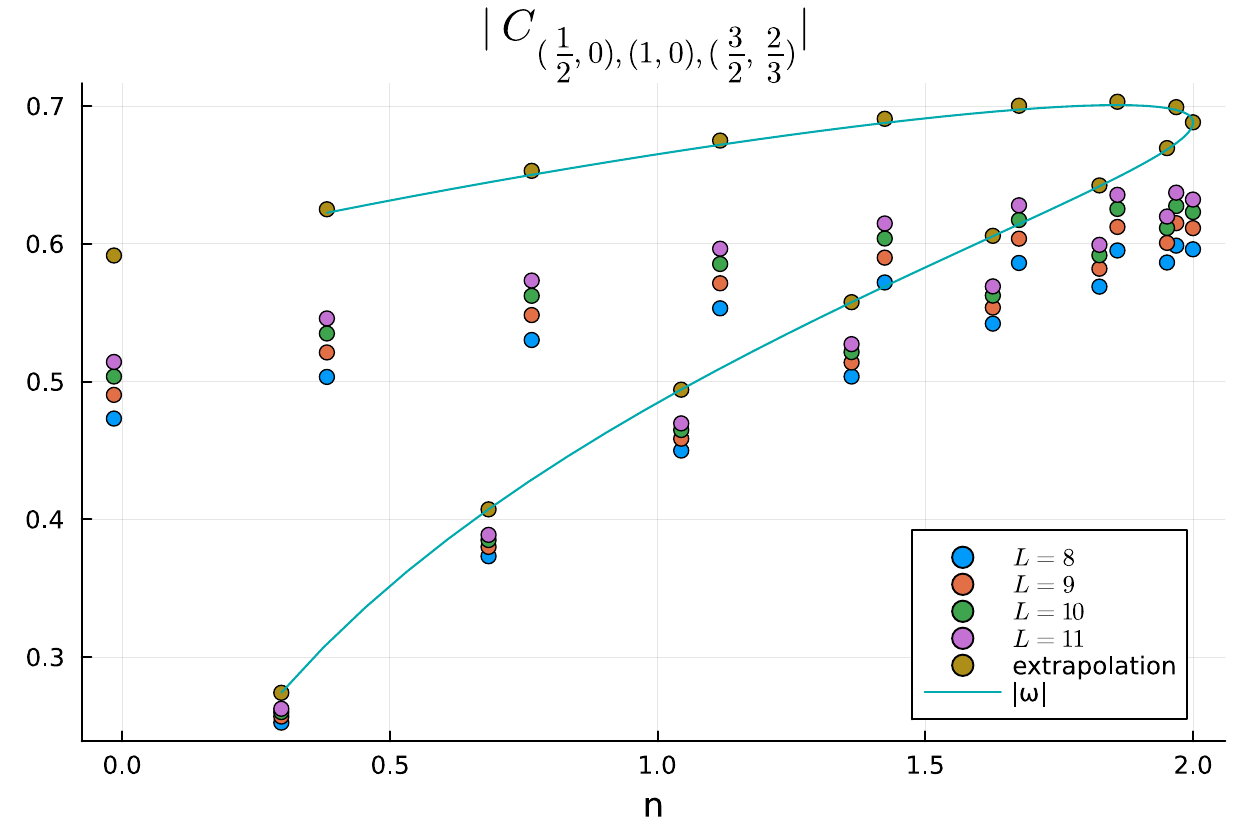} \quad
  \includegraphics[width=0.45\textwidth]{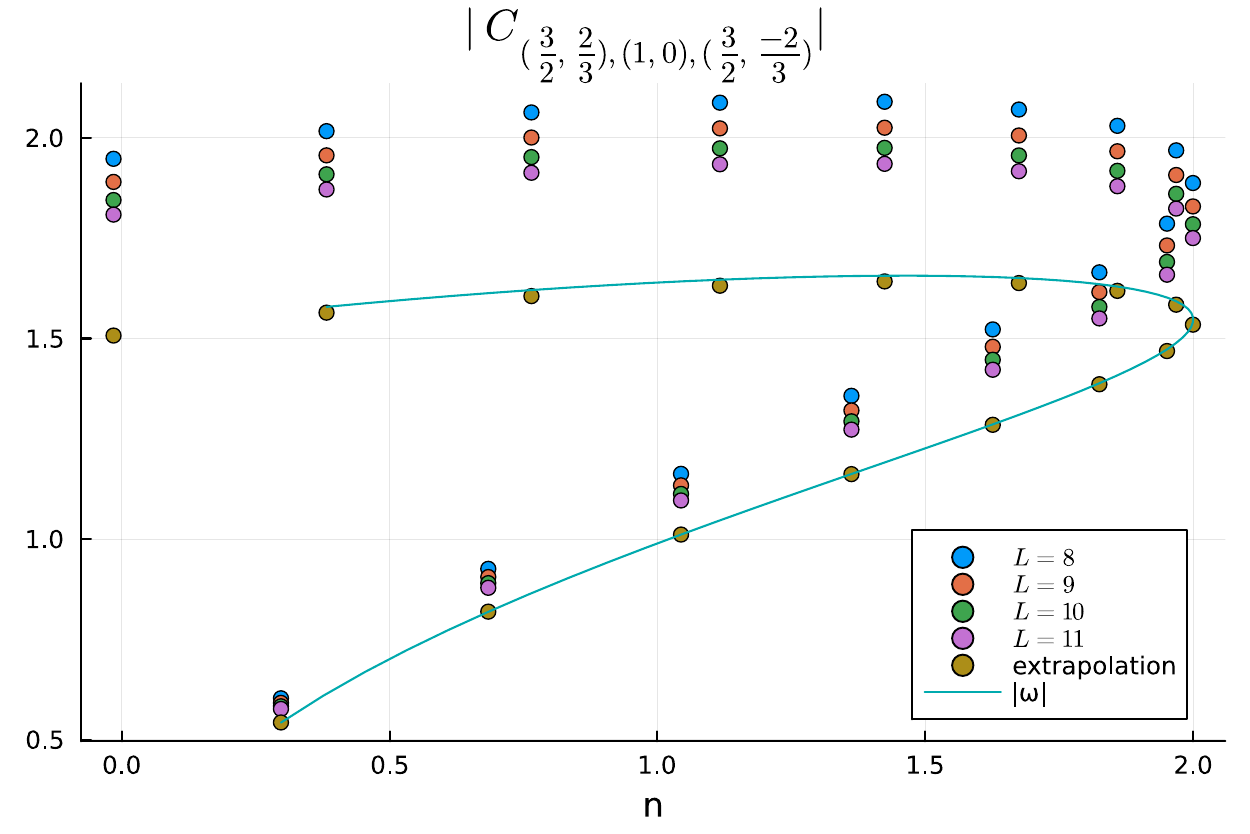}
  \medskip                  
  \includegraphics[width=0.45\textwidth]{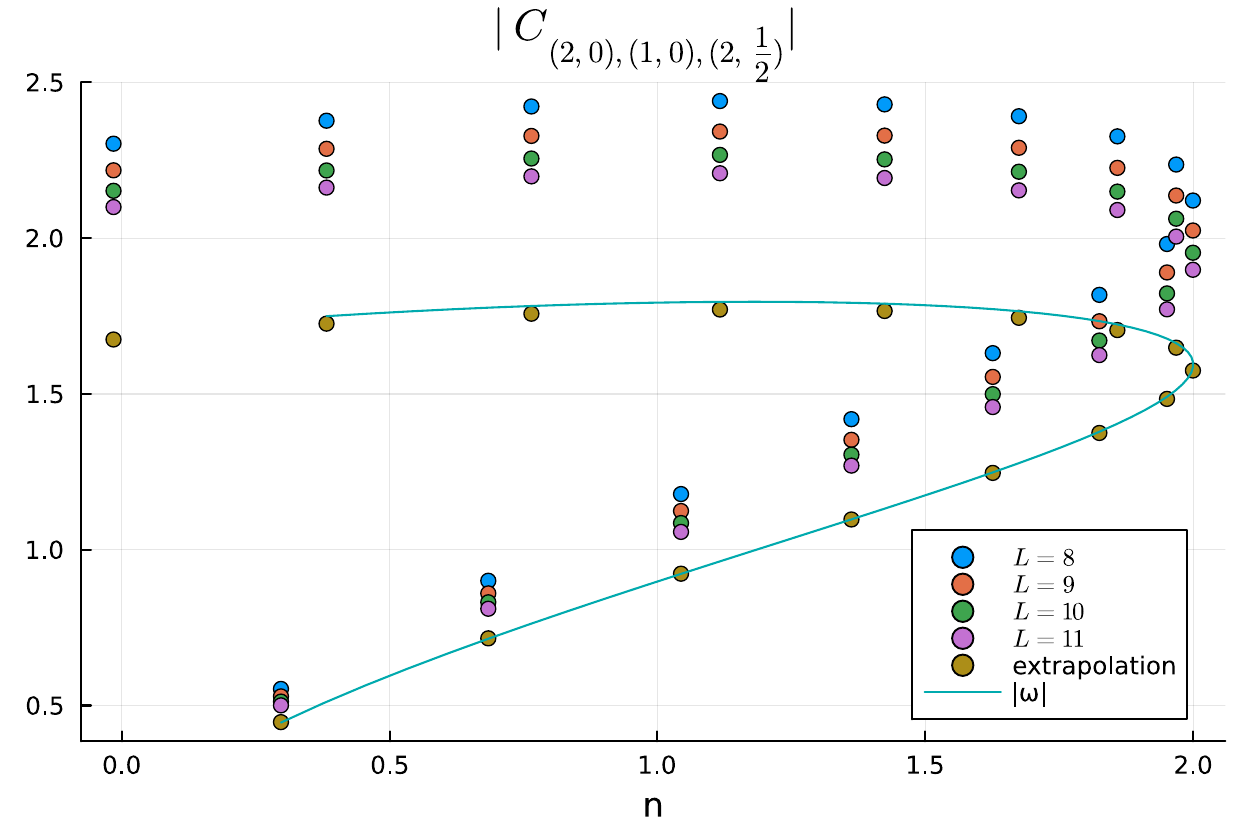} \quad
  \includegraphics[width=0.45\textwidth]{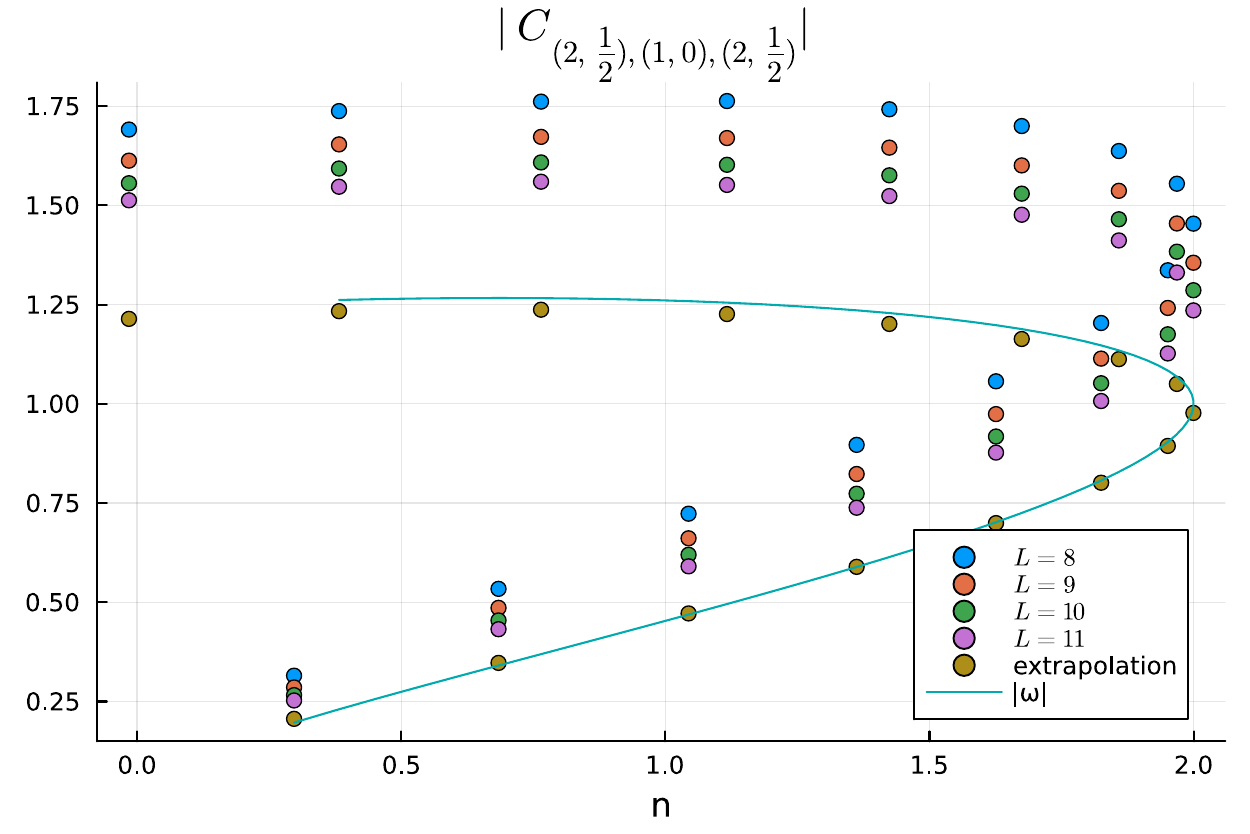} \quad
  \caption{Plots of $|C_{123}(L)|$ for the $O(n)$ loop-model results, their $L\to \infty$ extrapolations, and $|\omega_{(\frac12, 0),(1, 0),(\frac32, \frac23)}|$, $|\omega_{(\frac32,\frac23), (1, 0), (\frac32, -\frac{2}3)}|$ (top row, from left to right), $|\omega_{(2, 0), (1, 0), (2, \frac12)}|$, $|\omega_{(2, \frac12),(1, 0),(2, \frac12)}|$ (bottom row).}
  \label{fig:3pt_spin_plots}
\end{figure}

Of the 43 cases tested, 12 have all $s_i \neq 1$:
\begin{align}
\label{eq:32}
\renewcommand{\arraystretch}{1.2}
\begin{array}{| c | c |}
  \hline
(r_1, r_2, r_3) & (s_1, s_2, s_3) \\
  \hline
(\frac12, 1, \frac32) & (0, 0, \frac23) \\
  \hline
(\frac32, 1, \frac32) & (0, 0, \frac23), (\frac23, 0, \pm\frac23) \\
  \hline
(2, \frac12, \frac32) & (\frac12, 0, 0), (0, 0, \frac23), (\frac12, 0, \pm\frac12) \\
  \hline
(2, 1, 1)             & (\frac12, 0, 0) \\
  \hline
(2, 1, 2)             &(0, 0, \frac12), (\frac12, 0, \pm\frac12) \\
  \hline
\end{array}
\end{align}
In all cases, the algorithm described in section \ref{subsec:3ptspin} successfully gives results that converge to the expected $\omega_{123}$.
Figure \eqref{fig:3pt_spin_plots} shows a few examples, for which we plot the results in absolute value simply to avoid having to resolve the phase ambiguity in the definition of $d_{0, 0, 0}$. The rest of our numerical results can be found in the repository \cite{roux25}.

\subsection{Cases involving $V_{(r, 1)}$}\label{subsec:case_s=1}

\subsubsection{Degeneracy of the ground state}

In the CFT at generic central charge, the primary fields $V_{(r,s)}$ and $V_{(r,-s)}$ each generate an irreducible representation of the conformal algebra if $s\in [0, 1)$, but they belong to the same  indecomposable representation if $s=1$ \cite{nr20}. This peculiar feature of the case $s=1$ also occurs in the lattice model.
To see this, let us introduce the space parity $P$ which reflects link patterns with respect to the vertical axis, and the lattice translation operator $u$ defined by the diagram
\begin{align}
  \label{eq:22}
  u =\hspace{0.5cm} \begin{tikzpicture}[scale=0.5, baseline=(current bounding box.center)]
    \foreach \i in {0, 1, 2, 4, 5}
    \draw[fill] (\i, 2) circle (2pt);
    \foreach \i in {0, 1, 3, 4} {
      \draw[fill] (\i, 0) circle (2pt);
      \draw[thick, blue] (\i, 0) to[out=90, in=-90] ++(1, 2);
    }
    \draw[fill] (5, 0) circle (2pt);
    \draw[thick, blue] (5, 0) to[out=90, in=-120] ++(0.5, 1);
    \draw[thick, blue] (-0.5, 1) to[out=60, in=-90] ++(0.5, 1);
    \draw[dashed, black!50] (-0.5, 0) -- (-0.5, 2);
    \draw[dashed, black!50] (5.5, 0) -- (5.5, 2);
  \end{tikzpicture}\hspace{0.3cm},
\end{align}
The action of the transfer matrix on the modules $\mathcal{W}_{(r, 1)}$ is parity and translation invariant:
\begin{align}
    [T, u] = 0\quad , \quad
    [T, P] = 0\quad , \quad
    Pu = u^{-1}P\ .
\end{align}
The operators $T$ and $u$ can be simultaneously diagonalised. Because $T$ is parity invariant, $P$ preserves eigenspaces of $T$, while it reverses the lattice momentum. Since we know there exists a vector $| V_{(r, 1)} \rangle$ in the leading eigenspace of $T$ with $u^2$-eigenvalue $e^{4 i \pi / L}$, $P | V_{(r, 1)} \rangle$ is a distinct eigenvector with the same eigenvalue for $T$.
Note that we use the operator $u^2$ here because subtleties arise when comparing eigenvalues of $u$ with the conformal spin \cite{cs23}.
In contrast, the action of $T$ on $\mathcal{W}_{(r, s\neq 1)}$ is not parity invariant: it maps $\mathcal{W}_{(r, s)} \to \mathcal{W}_{(r, -s)}$, since parity reverses the orientations of the cyclic permutations.
Since our algorithm of section \ref{subsec:3ptspin} does not select either of the leading eigenvectors, the argument of section \ref{subsec:limit} does not apply and there is no guarantee that the limit exists.

We could in principle avoid this issue by computing the leading eigenvectors of $T$. Our approach is instead to rely on $T^M$ to be dominated by the ground state at large $M$, which is cheaper numerically, but can lead to uncontrolled effects when $T$ has 2 ground states.

\subsubsection{Numerical results}

Despite the problem of the ground state degeneracy, we experimentally find that in many cases, introducing simple factors of $\frac{L}{\pi}$ and $\sqrt2$ in \eqref{eq:21} is enough to obtain excellent convergence to $\omega_{123}$, or to combinations of $\omega_{123}$ with its images under parity transformations. The table below summarises the numerical results that we obtained. The limit is obtained from equation \eqref{eq:21}. Every time we indicate $0$ as the limit, in finite size $C_{123}(L) \ll 1$, which means that our argument of section \ref{subsec:limit} does not apply. An empty limit indicates that our numerical data are insufficient to reliably extrapolate the limit.

\begin{align}
\renewcommand{\arraystretch}{1.2}
\begin{array}{| c | c | c | c | c |}
  \hline
  (r_1, r_2, r_3) & (s_1, s_2, s_3) & \alpha_{123} & f_{123}(L) & \text{limit} \\
  \hline
  \multirow{5}{*}{$(1, 1, 1)$}            & (0, 0, 1) & -1  & \frac{1}{\sqrt{2}} & \omega_{(1,0)(1,0)(1,1)} \\
   & (0, 1, 0) & 1 & \frac{1}{\sqrt{2}}  & \omega_{(1,0)(1,1)(1,0)}   \\
              & (0, 1, 1) & 0  & 1                    & \omega_{(1, 0)(1, 1)(1, 1)} + \omega_{(1, 0)(1, 1)(1, -1)} \\
              & (1, 0, 1) & 0  & 1                    & \omega_{(1, 0)(1, 1)(1, 1)} - \omega_{(1, 0)(1, 1)(1, -1)} \\
              & (1, 1, 1) & 1 & \sqrt{2}  & \omega_{(1, 1)(1, 1)(1, 1)} - \omega_{(1, 1)(1, 1)(1, -1)} \\
  \hline
  \multirow{2}{*}{$(\frac12, 1, \frac32)$} & (0, 1, 0)           & &  & 0 \\
                                & (0, 1, \frac{2}{3}) & 1 & \sqrt{2} & \omega_{(\frac12, 0)(1, 1)(\frac32, \frac23)} - \omega_{(\frac12, 0)(1, -1)(\frac32, \frac23)} \\
  \hline
  \multirow{4}{*}{$(\frac{3}{2}, 1, \frac{3}{2})$} & (0, 1, 0)                   & & & 0 \\
                                  & (0, 1, \frac23)                  &1 & \sqrt2 & \omega_{(\frac32, 0)(1,1)(\frac32,\frac23)} + \omega_{(\frac32, 0)(1,-1)(\frac32,\frac23)} \\
                                  & (\frac23, 1, \frac23)                & 1   & \sqrt2 & \omega_{(\frac32, \frac23)(1,1)(\frac32,\frac23)} + \omega_{(\frac32, \frac23)(1,-1)(\frac32,\frac23)} \\
                                & (\frac23, 1, -\frac23)               & 1    & \sqrt2 & \omega_{(\frac32, \frac23)(1,1)(\frac32,-\frac23)} + \omega_{(\frac32, \frac23)(1,-1)(\frac32,-\frac23)} \\
  \hline
\multirow{6}{*}{$(2, 1, 2)$} & (0, 1, 0) &  1 & \frac{1}{\sqrt2} & \omega_{(2,0)(1,1)(2,0)} \\
              & (0, 1, \frac12)       & 1 & \sqrt2 & \omega_{(2,0)(1,1)(2,\frac12)} \\
              & (0, 1, 1)             &  & &  \\
              & (\frac12, 1, \frac12) & 1 & \frac{1}{\sqrt2} & \omega_{(2,\frac12)(1,1)(2,\frac12)} \\
              & (\frac12, 1, 1)       &  & &   \\
              & (\frac12, 1, -\frac12)&  &  &  \\
  \hline
  \multirow{6}{*}{$(2, 1, 1)$} & (0, 1, 0)           &                  & &  0 \\
              & (0, 1, 1)           & 0 & 1              & \omega_{(2, 0)(1, 1)(1, 1)} - \omega_{(2, 0)(1, 1)(1, -1)} \\
              & (\frac12, 1, 0)     & 1 & \sqrt2 & \omega_{(2, \frac12)(1, 1)(1, 0)} - \omega_{(2, \frac12)(1, -1)(1, 0)} \\
              & (\frac12, 1, 1)     &  &                &  \\
              & (1, 1, 0)           &   &            &  \\
              & (1, 1, 1)           &  &              &  \\
  \hline
  \multirow{2}{*}{$(2, \frac12, \frac23)$} & (1,0,0) &              &                  &  \\
              & (1,0,\frac23)        &  &                &  \\
  \hline
\end{array}
\label{tab}
\end{align}

\section{Outstanding issues}

\subsubsection*{Comparison with the probabilistic approach}

Using conformal loop ensembles, probability theorists have recently proved~\eqref{wott} in the case of three diagonal fields \cite[2021 version]{acsw21} and for three spinless 2-leg fields \cite[2024 version]{acsw21}. It would clearly be of interest to prove the general case, but it is our understanding from private communications that the probability-theory arguments presently have to be made on a case-by-case basis. It is also possible that probability theory can shed more light on the validity of our lattice approach.

\subsubsection*{Relation with integrable models}

The expression~\eqref{eq:C123spin} for the structure constant provides a potentially fruitful link to the theory of integrable models. It contains a ratio of two form factors times some additional factors that express the overlap of specific boundary states with left and right eigenvectors.
Form factors and correlation function of local fields have been thoroughly studied by
advanced Bethe Ansatz techniques (see \cite{jms21} for a review). Most of these studies focus on the 6-vertex model
(or XXZ spin chain in the Hamiltonian version) which is related to our $PSU(n)$ loop model, whereas the $O(n)$ model used in most of this work calls for
the slightly less studied 19-vertex model (or Izergin-Korepin $a_2^{(2)}$ spin chain).

We see several challenges for computing the form factors in~\eqref{eq:C123} by quantum integrability techniques. First, form factors have mainly been studied between the ground state and an excited state, whereas the form factor in the numerator of~\eqref{eq:C123} is between two excited states. Second, the operator ${\mathcal O}_2$ has some inherent non-local features, and it is not clear if it can be related to the local operators (products of Pauli matrices) usually studied. Third, and presumably most importantly, the change of representation from Temperley-Lieb to the spin chain may lead to a modification of the result, as witnessed by comparing existing computations of the emptiness-formation probability in the XXZ vertex \cite{bks03,stephan14} and $PSU(n)$ loop \cite[Section 7]{mn04} representations. Also the overlap scalar products in~\eqref{eq:C123spin} present noticeable differences with similar quantities previously considered \cite{djs10,gjp16}.

\subsubsection*{State-operator correspondence and fusion in the periodic Temperley-Lieb algebra}

Due to the non-local nature of loops, the construction of the operator $\mathcal{O}_{(r, s)}$ \eqref{eq:26}in Sections \ref{subsec:3ptspinless} and \ref{subsec:3ptspin} required generalising the usual state-operator correspondence by introducing link patterns with additional labels.
The recent construction of a fusion product for the periodic Temperley-Lieb algebra in \cite{im25} provides an alternative way to construct states in $\mathcal{W}_{(r_3, s_3)}$ from the data of two modules, via
\begin{align}
\label{eq:33}
\mathcal{W}_{(r_1, s_1)} \times_f \mathcal{W}_{(r_2, s_2)} \to \mathcal{W}_{(r_3, s_3)}.
\end{align}
It would be interesting to understand the relation between this construction and our morphism $\mathcal{O}_{(r, s)}$.

\subsubsection*{Uniform spanning trees}

Our conjecture \eqref{wott} provides non-trivial predictions in the limit $\beta^2 \to \frac12$ of uniform spanning trees,
a model which has been thoroughly investigated by combinatorialists (see e.g.\ \cite{kw11}).
In this case, the result $n^{-\frac12} \omega_{(1,0)(1,0)(1,0)} = 1$ is trivial: any set of two or three points belong to the same loop,
since there is only one loop in the problem (the one which separates the tree from its dual).

\hspace{-16pt}
\begin{minipage}{0.6\textwidth}
  \setlength{\parindent}{1.5em}
  However, a result like $n^{-\frac12} \omega_{(2,0)(1,0)(1,0)} = 0.819035153\cdots$ is clearly less trivial.
  The factor $C_{123}$ is here related to the probability that one point is at the junction of two long branches in the tree,
  each of which extends to one of the two other points; see Figure \eqref{fig:tree}.
This is a very unlikely situation in a uniform spanning tree, but it is normalised to a finite number
by the 2-point functions entering in~\eqref{wott}, some of which are now non-trivial. 
Similarly, cases of \eqref{ustlim} with $r \neq 1$ are non-trivial, notwithstanding the simple value $n^{-\frac12} \omega_{(r,0)(r,0)(1,0)} = 1$.
For $r \in \mathbb{N}$, they can be related to situations where long branches of the uniform spanning tree join or touch in various ways, as can be seen from drawing.
\end{minipage}
\begin{minipage}{0.35\textwidth}
  \begin{align}
  \hspace{-0.5cm}
    \begin{tikzpicture}[
      baseline = (current bounding box.center)
      scale=1.2,
      vertex/.style={circle, draw=black, fill=red!70, inner sep=1pt},
      marked/.style={draw=black, line width=2pt, cross out, minimum size=6pt, inner sep=0pt},
      tree/.style={line width=1.2pt, blue, line cap=round},
      grid/.style={gray!40, line width=0.5pt},
      loop/.style={line width=2pt, red!50, rounded corners=10pt}]
      \draw[grid] (0,0) grid (3,3);
      \foreach \x in {0,1,2,3} {
        \foreach \y in {0,1,2,3} {
          \node[vertex] at (\x,\y) {};
        }
      }
      \draw[tree] (0,3) -- (0,2) -- (0,1) -- (1,1) -- (1,3);
      \draw[tree] (1,1) -- (2,1) -- (3,1) -- (3,0) -- (2, 0);
      \draw[tree] (2,1) -- (2,2) -- (3,2);
      \draw[tree] (3,3) -- (2,3) -- (2,2);
      \draw[tree] (1,0) -- (0,0) -- (0,1);
      \draw[loop] (1.5, 1) -- (1, 1.5) -- (1.5, 2) -- (1, 2.5) --
      (1.5, 3) -- (1, 3.5) -- (0.5, 3) -- (1, 2.5) -- (0.5, 2) -- (1, 1.5)
      -- (0.5, 1) -- (0, 1.5) -- (0.5, 2) -- (0, 2.5) -- (0.5, 3) -- (0, 3.5)
      -- (-0.5, 3) -- (0, 2.5) -- (-0.5, 2) -- (0, 1.5) -- (-0.5, 1) -- (0, 0.5) -- (-0.5, 0)
      -- (0, -0.5) -- (0.5, 0) -- (1, -0.5) -- (1.5, 0) -- (1, 0.5) -- (0.5, 0) -- (0, 0.5)
      -- (0.5, 1) -- (1, 0.5) -- (1.5, 1) -- (2, 0.5) -- (2.5, 1) -- (3, 0.5) -- (2.5, 0)
      -- (2, 0.5) -- (1.5, 0) -- (2, -0.5) -- (2.5, 0) -- (3, -0.5) -- (3.5, 0) -- (3, 0.5)
      -- (3.5, 1) -- (3, 1.5) -- (2.5, 1) -- (2, 1.5) -- (2.5, 2) -- (3, 1.5) -- (3.5, 2) -- (3, 2.5)
      -- (2.5, 2) -- (2, 2.5) -- (2.5, 3) -- (3, 2.5) -- (3.5, 3) -- (3, 3.5) -- (2.5, 3) -- (2, 3.5)
      -- (1.5, 3) -- (2, 2.5) -- (1.5, 2) -- (2, 1.5) -- cycle;
      \node[marked] at (1.5, 1) {};
      \node[marked] at (0, 3) {};
      \node[marked] at (2, 0) {};
    \end{tikzpicture}
  \label{fig:tree}
  \end{align}
\end{minipage}
\subsubsection*{Comparison with higher-dimensional CFT}

Our transfer matrix method uses standard ideas for approximating 2d CFT. The difficulties that we faced arose from adapting the method to loop models, and dealing with the operators' legs. The use of similar methods in higher-dimensional CFT is more recent, and relies on the nontrivial idea of the fuzzy sphere regularization. This idea can be applied to computing 3-point structure constants \cite{akty25}. It would be interesting to compare the behaviour of lattice 3-point functions in 2d with their behaviour in higher dimensions: in particular, how to deal with nonzero spins (which we found tricky in loop models), and the dependence on the lattice size.

\section*{Acknowledgements}

We are grateful to Xin Sun for communicating results of \cite[2024 version]{acsw21} before publication, and for inspiring discussions.
This has stimulated us to investigate three-point functions, and helped us find the right analytic ansatz.

We thank Max Downing for suggesting several clarifications after reading a draft version of this paper.

We are grateful to Hubert Saleur for pointing out subtleties that arise when comparing eigenvalues of the translation operator $u$ to the conformal spin, and for urging us to better understand the case of operators with spin.

JJ thanks Yacine Ikhlef and Hubert Saleur for a collaboration in 2013, during which some of the numerical results presented here in the first part of Section~\ref{sec:res} were obtained, and which eventually led to the publication \cite{ijs15}. He also thanks Nathanaël Berestycki for a discussion on uniform spanning trees.

We acknowledge funding from the Agence Nationale de la Recherche through the grant CONFICA (grant No.\ ANR-21-CE40-0003) and from the European Research Council through the ERC-SyG project ReNewQuantum.

\bibliographystyle{../../inputs/morder8}
\bibliography{../../inputs/992.bib}
\end{document}